\documentclass[letterpaper,english,reprint, aps]{revtex4-1}
\usepackage[T1]{fontenc}
\usepackage[latin9]{inputenc}
\usepackage{color}
\usepackage{verbatim}
\usepackage{textcomp}
\usepackage{multirow}
\usepackage{amsmath}
\usepackage{amssymb}
\usepackage{graphicx}
\usepackage{hyperref}
\PassOptionsToPackage{normalem}{ulem}
\usepackage{ulem}

\makeatletter

\pdfpageheight\paperheight
\pdfpagewidth\paperwidth

\newcommand{\lyxmathsym}[1]{\ifmmode\begingroup\def\b@ld{bold}
  \text{\ifx\math@version\b@ld\bfseries\fi#1}\endgroup\else#1\fi}

\providecommand{\tabularnewline}{\\}

\newcommand{\COMMENTED}[1]{}

\newcommand{\CHANGED}[1]{#1}
\DeclareMathOperator*{\argmax}{arg\,max}

\makeatother

\usepackage{babel}
\begin{document}
\preprint{APS/123-QED}
\title{A structural optimization algorithm with stochastic forces and stresses}
\author{Siyuan Chen}
\email{schen24@email.wm.edu}

\affiliation{Department of Physics, College of William \& Mary, Williamsburg, VA}
\author{Shiwei Zhang}
\email{szhang@flatironinstitute.org}

\affiliation{Center for Computational Quantum Physics, Flatiron Institute, New
York, NY}
\begin{abstract}
We propose an %
 algorithm for optimizations in which %
 the gradients
contain stochastic noise. This arises, for example, in structural optimizations
when computations of forces and stresses rely on methods involving
Monte Carlo sampling, such as quantum Monte Carlo or neural network
states, %
or are performed on quantum devices which
have intrinsic noise. Our proposed algorithm is based on the combination
of two key ingredients: an update rule derived from the steepest descent
method, and a staged scheduling of the targeted statistical error
and step-size, with position averaging. We compare it with commonly
applied algorithms, including some of the latest machine learning
optimization methods, and show that the algorithm consistently performs
efficiently and robustly under realistic conditions. 
Applying this algorithm, we achieve 
full-degree optimizations in solids using \textit{ab initio} many-body computations, by auxiliary-field
quantum Monte Carlo with planewaves and pseudopotentials. A new metastable
structure in Si was discovered in a mixed geometry and lattice relaxing
simulation. In addition to structural optimization in materials, our
algorithm can potentially be useful in other problems in various fields where optimization
with noisy gradients is needed.
\end{abstract}

\maketitle

\section{Motivation\label{sec:motivation}}

Geometry optimization is the procedure to locate the structure with
energy or free-energy minimum in a solid or molecular system given
the atomic compositions. Such a local or global minimum state is usually
a naturally existing structure under common or extreme conditions.
As an essential ingredient in materials discovery and design, structural
search and geometry optimization have important applications
from quantum materials to catalysis to protein folding to drug design,
covering wide-ranging areas including condensed matter physics, materials
science, chemistry, biology, etc. The problems involved are fundamental,
connecting applied mathematics, algorithms, and computing with quantum
chemistry and physics. With the rapid advent of computational methods
and computing platforms, they have become a growing component of the
scientific research repertoire, complementing and in some cases supplementing
experimental efforts. 

The vast majority of geometry optimization efforts to date have been
performed with an effective ion-ion potential 
(force fields) \cite{Frenkel_FF_2007,Leach_FF_2001}, or \textit{ab initio} molecular dynamics 
based on density-functional theory
(DFT) \cite{Hohenberg_PR_1964,Jones_RMP_2015,Becke_JCP_2014,Burke_JCP_2012}. 
Force fields are obtained empirically from experimental data, or derived 
from DFT calculations at fixed structures, or learned from combinations of theoretical or experimental data.
Geometry optimization using force fields is computationally low-cost %
and convenient, and
allows a variety of realistic calculations to be performed.
The development of \textit{ab initio}
molecular dynamics \cite{CarParrinello_PRL_1985} signaled a fundamental step forward in accuracy and predictive power, where the interatomic
forces are obtained more accurately from DFT on the fly, allowing the structural optimization to better capture the underlying 
quantum mechanical nature.
With either force fields or \textit{ab initio} DFT, the total energy and forces can be obtained deterministically
\CHANGED{without any statistical noise,} %
and a well tested set of optimization procedures have
been developed and applied. 

In many quantum materials, however, Kohn-Sham DFT is still not sufficiently accurate,
because of its underlying independent-electron framework,
and a more advanced
treatment of electronic correlations is needed to provide reliable structural predictions.
Examples of such materials include %
so-called strongly correlated systems, which encompass a broad range of materials
with great fundamental and technological importance.  One of
the frontiers in quantum science is to develop computational methods
which can go beyond DFT-based methods in accuracy, with reasonable
computational cost. Progress has been made from several fronts, for
example, with the combination of DFT and the GW \cite{Hedin_PR_1965}, approaches
based on dynamical mean field theory (DMFT) \cite{Georges_RMP_1996},
quantum Monte Carlo \CHANGED{
(QMC) methods \cite{Foulkes_RMP_2001,Zhang_PRL_2003,Rillo_JCP_2018,Tirelli_arXiv_2021},
}
quantum chemistry methods \cite{Levine_QChem_1991,Cramer_CChem_2002},
etc. For instance, the computation of
forces and stresses with plane-wave auxiliary field quantum Monte
Carlo (PW-AFQMC) \cite{Zhang_PRL_2003,Suewattana_PRB_2007} has recently been demonstrated \cite{SC_fs_paper},
paving the way for  \textit{ab initio}  geometry optimization in this  many-body
framework. 

One crucial new aspect of geometry optimization with most of the post-DFT methods
is that \CHANGED{information of the potential energy surface (PES)} %
obtained from such approaches contain statistical uncertainties. 
The post-DFT methods, because of the exponential scaling of the
Hilbert space in a many-body treatment, often involve stochastic
sampling. This includes the various classes of QMC
methods, but other approaches such as DMFT may also contain ingredients
which rely on Monte Carlo sampling. Neural network wave function approaches \cite{Jia_AQT_2019,Carleo_Science_2017} also typically involve stochastic ingredients. 
Additionally if the many-body computation is performed on a quantum device \cite{Lanyon_NatChem_2010,Huggins_Nature_2022} %
noise may also be present.

Geometry optimization under these situations, namely with %
noisy \CHANGED{PES information,} %
presents new challenges, and also new opportunities. As we illustrate below, the
presence of statistical noise 
\CHANGED{\sout{in the computed gradients}}  %
can fundamentally
change the behavior of the optimization algorithm. On the other hand,
the fact that the size of the statistical error bar can be controlled
by the amount of Monte Carlo sampling affords opportunities to tune
and adapt the algorithm to minimize the integrated computational cost
in the optimization process. \CHANGED{
To date work on structural optimization with noisy PES has not been widespread. One class \cite{Guareschi_JCTC_2013,Zen_JCTC_2012,Barborini_JCTC_2012}
focuses on applications using variational Monte Carlo (VMC), 
which allows computations of forces and Hessians besides the total energy, mostly applying standard optimization algorithms. 
Another class focuses on using total energies for exploring the PES \cite{Wagner_PRL_2010,Tiihonen_JCP_2022}, since computing forces and other gradients remains challenging in QMC, especially with projection methods beyond VMC. 
General and more systematic applications of structural optimization in correlated materials will in all likelihood require going beyond VMC, and effectively exploiting accurate forces and other gradients to efficiently scale up to high dimensions. 
The present work investigates optimization algorithms with this as the background.
}
In principle, a number of algorithms 
widely used in the machine-learning community can be adopted
to the geometry optimization problem.
However, we find that, in a variety of realistic situations under
general conditions, the performance of these algorithms is often sub-optimal.
Given that the many-body computational methods tend to have higher
computational costs, it is essential to minimize the number of
times that force or stress needs to be evaluated,
and the amount of sampling in each evaluation, before the optimized structure is reached.

In this paper, we propose an algorithm for optimization when the computed
gradients have intrinsic statistical noise. The algorithm is found
to consistently yield efficient and robust performance in geometry
optimization using stochastic forces and stresses, often outperforming
the best existing methods. We apply the method to realize a full geometry
optimization using forces and stresses computed from PW-AFQMC. In
analyzing and testing the method, we unexpectedly discovered a new
orthorhombic $Cmca$ structure in solid silicon. 

The rest of the paper is organized as follows. In Sec.~\ref{sec:overview} we give an overview  
of our algorithm and outline the two key components. This is 
followed by an %
analysis  in Sec.~\ref{sec:analysis}, with comparisons to common geometry optimization algorithms,
including leading machine learning algorithms. 
In Sec.~\ref{sec:results} we apply our method to perform, for the first time, a full geometry optimization in solids
using PW-AFQMC. We then describe %
the discovery of the
new structure in Si in Sec.~\ref{sec:results2}, before concluding in Sec.~\ref{sec:conclusion}.

\section{Algorithm Overview\label{sec:overview}}

A noisy gradient, such as an interatomic force evaluated from a QMC calculation,
can be written as
\begin{equation}
\tilde{\mathbf{F}}=\mathbf{F}+\vec{\varepsilon}\,,
\end{equation}
where $\mathbf{F}$ is the true force, and
$\tilde{\mathbf{F}}$ is the (expectation) value computed by the numerical method with stochastic
components. The vector $\vec{\varepsilon}$ %
denotes stochastic
noise, for example the statistical error bar estimated from the QMC
computation. In the case of a sufficiently large number of Monte Carlo
samples (realized in most cases but not always), the central limit
theorem dictates that the noise is given by a Gaussian
\begin{equation}
\varepsilon_{i}=\mathcal{N}(0,s_{i}^{2})\,,
\end{equation}
where $i$ denotes a component of the gradient
(e.g. a combination of the atom number and the Cartesian direction in the case of interatomic forces), and
$s_{i}\propto N_{s}^{-1/2}$ is the standard deviation,
which can be reduced as the square-root of the number of effective
samples $N_{s}$.
The computational cost is typically proportional to $N_{s}$.

Our algorithm consists of two key components. Inside each step of
the optimization, we follow an update rule using the current 
$\tilde{\mathbf{F}}$, %
which is a fixed-step-size modification of 
\CHANGED{the gradient descent with momentum method \cite{Qian_momentum_1999,Rumelhart_momentum_1986}, which we will refer to as}
\textquotedblleft fixed-step steepest descent\textquotedblright{}
(FSSD). Globally, the optimization process is divided into stages,
each with a 
\CHANGED{target statistical error}
$s$ for $\vec{\varepsilon}$
(hence controlling the computational cost per gradient evaluation)
and specific choice of step size, called a staged error-targeting
(SET) workflow. The SET is complemented by a self-averaging procedure
within each stage which further accelerates convergence. We outline
the two ingredients separately below, and provide analysis and discussions
in the following sections.

\subsection{The FSSD update rule\label{subsec:FSSD}}

The SET approach discussed in Sec.~\ref{subsec:SET} defines the overall
algorithm. Each step inside each stage of SET is taken with the FSSD
algorithm, which works as follows. Let $n$ denote the current step
number, and $\mathbf{x}_{n}$ denote the atom positions at the end
of this step. Here, $\mathbf{x}_{n}$ is an $N_{d}$-dimensional vector,
with $N_{d}$ being the degree of freedom in the optimization.

(1) Calculate the force at the atomic configuration from
the previous step: $\mathbf{F}_{n-1}=-\nabla\ell(\mathbf{x}_{n-1})$.
(In the case of quantum many-body computations, the loss function
$\ell$ is the ground-state energy $E$, and the force is typically computed as
the estimator of an observable directly, for example via the the
Hellmann-Feynman theorem \cite{SC_fs_paper}.)

(2) The search direction is then chosen as
\begin{equation}
\mathbf{d}_{n}=\frac{\alpha\mathbf{d}_{n-1}+\mathbf{F}_{n-1}}{\alpha+1}\,,
\end{equation}

\noindent where $\mathbf{d}_{n-1}$ is the the displacement \textit{direction}
of the step $(n-1)$, which encodes the forces from past steps and
thus serves as a ``historic force.'' We experiment with the choice
of the parameter $\alpha$ (see Appendix \ref{sec:method-params}),
but typically set it to $\alpha=1/e$.

(3) The displacement \textit{vector} is now set to the chosen direction
from (2), with step size $L$ which is fixed throughout the stage:
\begin{equation}
\Delta\mathbf{x}_{n}=L\frac{\mathbf{d}_{n}}{|\mathbf{d}_{n}|}\,.
\end{equation}

(4) Obtain the new atom position vector, $\mathbf{x}_{n}=\mathbf{x}_{n-1}+\Delta\mathbf{x}_{n}$.
Account for symmetries and constraints such as periodic boundary conditions
or restricting degree of freedoms as needed.

\subsection{SET scheduling approach\label{subsec:SET}}

The staged error-targeting workflow (SET) can be described as follows:

(1a) Initialize the stage. At the beginning of each stage of SET,
the step count $n$ is set to 1, and an initial position $\mathbf{x}_{0}$
is given, which is either the input at the beginning of the optimization
or inherited from the previous stage {[}see (5) below{]}. We also
set $\mathbf{d}_{0}=0$ in (2) of Sec.~\ref{subsec:FSSD} (thus the
first step within each stage is a standard steepest descent).

(1b) Use a fixed step size $L$, and target a fixed
average statistical error bar $s$  for the force
computation throughout this stage. 
The values
of $L$ and $s$ are either input (first stage at the beginning of
the optimization) or set at the end of the previous stage {[}see (5)
below{]}.
 From $s$ we obtain an estimate
of the %
computational resources needed, $C(s)\propto s^{-2}$ for each force evaluation,
which helps to set the run parameters  during this stage (e.g.~population size and projection
time in AFQMC). 
We have used the average $\sum_{ia}s_{ia}/N_{d}$ for $s$, but clearly other choices are possible.
 To initialize the optimization we have typically used $L=0.1\sqrt{N_{d}}\,\mathrm{[Bohr]}$.
For $s$, we have
typically used an initial value of $\sim$20\% of
the average of each component of the initial force. These choices
are \textit{ad hoc} and can be replaced by other input values, for example, from an estimate
by a less computationally costly approach such as DFT.

(2) Do a step of FSSD with the current step-size $L$ and the rationed
computational resources $C(s)$. This consists of the steps described
in Sec.~\ref{subsec:FSSD}.

(3) Perform convergence analysis if a threshold number of steps have
been reached. Our detailed convergence analysis algorithm is discussed
in Appendix \ref{sec:conv-analysis-algo}.

(4) If the convergence is not reached in (3), loop back to (2) for
the next step within this stage; otherwise, the analysis will reveal
a \textit{previous} step count $m$ ($m<n$) where the convergence
was reached. Take the average of $\{\mathbf{x}_{m},\mathbf{x}_{m+1},\ldots,\mathbf{x}_{n}\}$
(see Appendix \ref{sec:conv-analysis-algo}) to obtain the final position
of this stage, $\bar{\mathbf{x}}$. 

(5) If overall objective of optimization is reached, stop; otherwise,
set $\mathbf{x}_{0}=\bar{\mathbf{x}}$, modify $L$ and $s$, and
return to (1). For the latter, we typically lower $s$ and $L$ by
the same ratio. %

\section{Algorithm Analysis\label{sec:analysis}}

In this section we analyze our algorithm, provide additional implementation
details, describe our test  setups,
and discuss additional algorithmic issues and further improvements. From the update-rule prospective,
we make a comparison in Sec.~\ref{subsec:FSSD-compare} between the FSSD and common line-search
\cite{RobbinsMonro_LS_1951,Armijo_LS_1966,Wolfe1,Wolfe2,Bertsekas_SIAMOpt_2006,Bertsekas_LS_2016}
based algorithms (steepest descent \cite{Debye_SD_1909} and conjugate gradient 
\cite{Magnus_CG_1952,Shewchuk_CG_1994,FletcherReeves_CG_1964,PolakRibiere_CG_1969,Schraudolph_CG_2003}), as well as several optimization algorithms widely
used in machine learning (RMSProp\cite{Tieleman_RMSProp_2012}, Adadelta\cite{Zeiler_Adadelta_2012}, and Adam\cite{KingmaBa_Adam_2014}).
Then in Sec.~\ref{subsec:sched-discuss} we analyze SET,  illustrate
how position averaging and staged scheduling %
improve %
the performance of the optimization
procedure, %
and discuss some potential improvements. %

To facilitate the study in this part, we create DFT-models to simulate actual many-body computations with noise. 
We consider a number of real solids and realistic geometry optimizations, but use forces and stresses computed from DFT, which is substantially less computationally costly than many-body methods.
Synthetic noise is introduced on the forces,
defining $\vec{\varepsilon}$ %
according to the targeted statistical errors of the many-body computation, and 
sampling $\tilde{\mathbf{F}}=\{ \tilde{F}_{i} \}$ from $\mathcal{N}(F_{i},s^{2})$, where $\{{F}_{i}\}$ are the corresponding forces or stresses computed from DFT.
As indicated above, we have chosen the noise to be isotropic in all directions based on
our observations from AFQMC, but this can be generalized as needed.
The DFT-model replaces the many-body computation, and is called to produce 
$\{ \tilde{F}_{i} \}$ %
as the input to the optimization algorithm. 
This provides a controlled, flexible, and convenient emulator for systematic studies of the performance of the optimization algorithm.

\subsection{FSSD vs.~line-search and ML algorithms\label{subsec:FSSD-compare}}

In the presence of noise in the gradients, standard line-search algorithms
such as steepest descent and conjugate gradient can suffer efficiency
loss or even fail to find the correct local minimum. (See Appendix
\ref{sec:fragile-line-search} for an illustration.) Many machine
learning (ML) methods, which avoid line-search and incorporate advanced
optimization algorithms for low-quality gradients, are an obvious
choice as an alternative in such situations. Our expectation was that
these would be the best option to serve as the engine in our optimization.
However, to our surprise we found that the FSSD was consistently competitive
with or even out-performed the ML algorithms in geometry optimizations in solids.

Below we describe two sets of tests in which we characterize the performance
of FSSD in comparison with other methods. For line-search methods,
we use the standard steepest descent, and the conjugate gradient with
a Polak-Ribi\'ere formula\cite{PolakRibiere_CG_1969}, which showed the best performance
within several conjugate gradient variants in our experiments. For
the ML algorithms we choose three: RMSProp, Adadelta, and Adam, which
are well-known and generally found to be among the best performing
methods for a variety of problems.
For each algorithm, we have experimented with the choice of step size or learning rate 
in order to choose an optimal setting for the comparison. 
(Details on the parameter choices can be
found in Appendix \ref{sec:method-params}.)

Figure \ref{fig:Si-compareLSML} shows a convergence analysis of
FSSD and other algorithms in solid Si (in which the targeted minimum is
the so-called $\beta$-tin structure, reached under a pressure-induced
phase transition from the diamond structure, as illustrated in the
top panel; see details in Appendix \ref{sec:structures}). Three random
runs are shown for each method. As seen in panel (b), the performance
of line-search methods, in which one line-search iteration can take
several steps, is lowered by the statistical noise. The convergence
of FSSD is not only much faster but also more robust than the two
line-search methods. The ML algorithms are shown in panel (c). RMSProp
shows slightly worse convergence speed and
quality than FSSD. 
These methods have conceptual similarities: both involve averaging over gradient history,
and both become %
a fixed-step approach %
when this averaging is turned off. Adadelta has excellent %
convergence quality, but slower convergence. 
Adam performs significantly worse %
than the other algorithms here.

\begin{figure}
\includegraphics[width=1\linewidth]{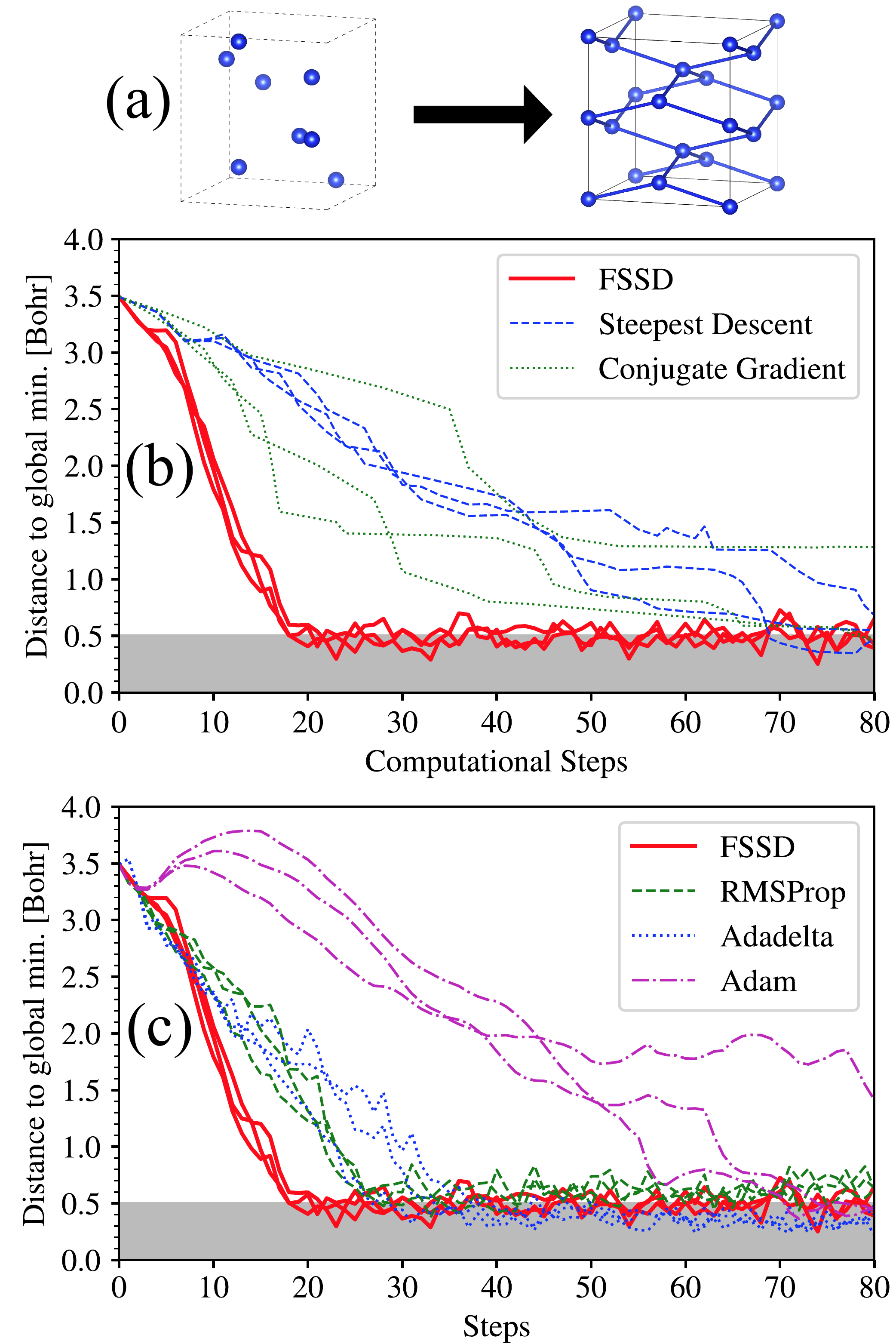}\caption{\label{fig:Si-compareLSML}Comparison of the convergence of fixed-step
steepest descent (FSSD) vs.~line-search and ML algorithms, %
in a problem of Si phase transition. Panel (a) shows the starting structure (50\% diamond:50\% $\beta$-tin),
compressed to the $\beta$-tin lattice constant, as well as the expected
final structure ($\beta$-tin), while panels (b) and (c) show the comparison of 
FSSD with line-search methods and ML methods, respectively.
In (b) and (c), distance to minimum is plotted vs.~number of computational steps. 
The gray region in (b) and (c) marks the ``convergence
region'' (defined as the Euclidean distance within 0.5 Bohr of the ideal $\beta$-tin final structure).
}
\end{figure}

We next compare FSSD and the three ML algorithms in a two-dimensional
solid, the $\mathrm{MoS_{2}}$ monolayer, which has an interesting energy landscape:
the global minimum (2H) and a nearby local minimum (1T) are separated
by a ridge, as depicted in Fig.~\ref{fig:MoS2-compareML} (system details in Appendix
\ref{sec:structures}). We observe
that the original ML algorithms all lead to the local minimum structure,
while FSSD finds the global minimum. We then modified the ML algorithms
and introduced a \textquotedblleft by-norm\textquotedblright{} variant
(details in Appendix \ref{sec:method-params}). As shown in Fig.~\ref{fig:MoS2-compareML}, %
this resulted in different behaviors
from the original \textquotedblleft element-wise\textquotedblright{}
algorithms, crossing over %
 the ridge and finding the global minimum instead.
These \textquotedblleft by-norm\textquotedblright{} algorithms, similar
to FSSD, follow paths that are almost perpendicular to the contour
lines, which lead to the global minimum in this setup.
\CHANGED{
It is worth emphasizing that the observation here should not be taken as a general conclusion over any energy landscape. The proximity of the initial structure to the convergence boundary is a key factor, }
but the markedly different behaviors from the different variants are still %
 interesting to note.

The convergence speed of each method in $\mathrm{MoS_{2}}$ can be
seen on the contour plot, where each arrow represents a single optimization
step; a more direct comparison is shown in panel (c). %
FSSD remains the fastest method, %
again closely followed by RMSProp and Adadelta. These tests
also confirm the characteristics of the ML algorithms seen in the
Si test: RMSProp is similar to FSSD, and shows relatively fast convergence
on the shortest route; %
Adadelta optimizes efficiently %
on steep surfaces but reduces the step size more drastically
when entering a \textquotedblleft flatter\textquotedblright{} landscape,
which %
slows down its final convergence; 
due to its inclusion of the first momentum Adam
produces a path that is more like a damping dynamics, %
delaying its convergence speed.

\begin{figure}
\includegraphics[width=1\linewidth]{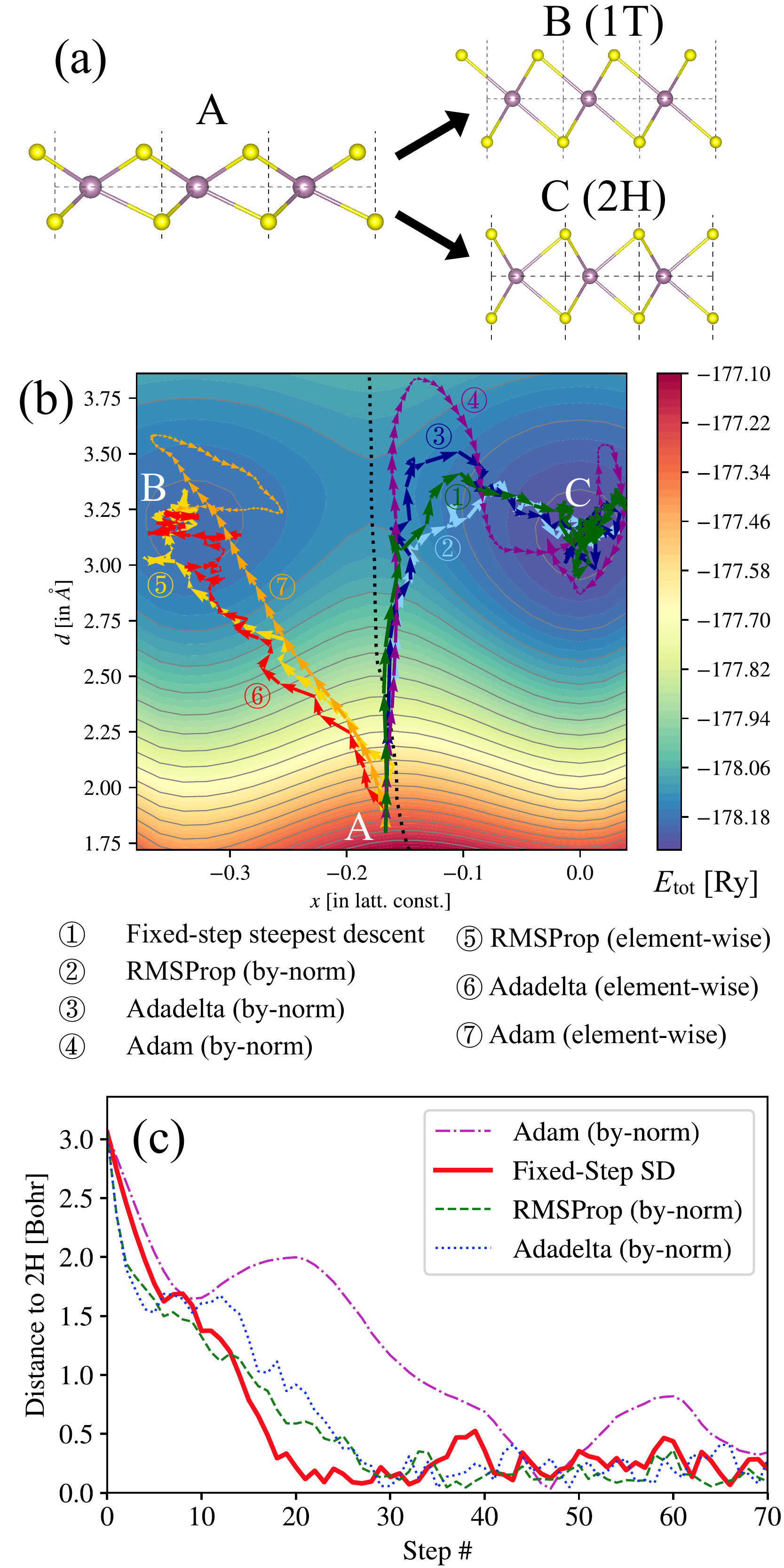}\caption{\label{fig:MoS2-compareML}Convergence and performance
in $\mathrm{MoS}_{2}$, for the FSSD and ML %
optimization algorithms. Panel (a) depicts the set-up:  
the initial geometry of compressed 50\% 1T:50\% 2H structure (A),
the 1T local minimum (B), and the 2H global minimum (C). 
Panel (b) shows 
the convergence trajectory %
of each algorithm in the $x$-$d$ plane (two of the nine degrees-of-freedom in the optimization, as defined
in Appendix \ref{sec:structures}, Fig.~\ref{fig:MoS2-setup}). The background color and contours
indicate the ground-state energy $E$ at each structure. Different colored curves with arrows
show the trajectories of 
different algorithms (labeled by numbers, as shown in the legends). The length of each arrow indicates the size
of the step in the $x$-$d$ plane. The black dotted curve marks the ``energy
barrier.'' %
Panel (c) compares the convergence speed
of FSSD and the ``by-norm'' variants of the ML algorithms. 
}
\end{figure}

\subsection{Performance and analysis of SET\label{subsec:sched-discuss}}

\begin{figure}
\includegraphics[width=1\linewidth]{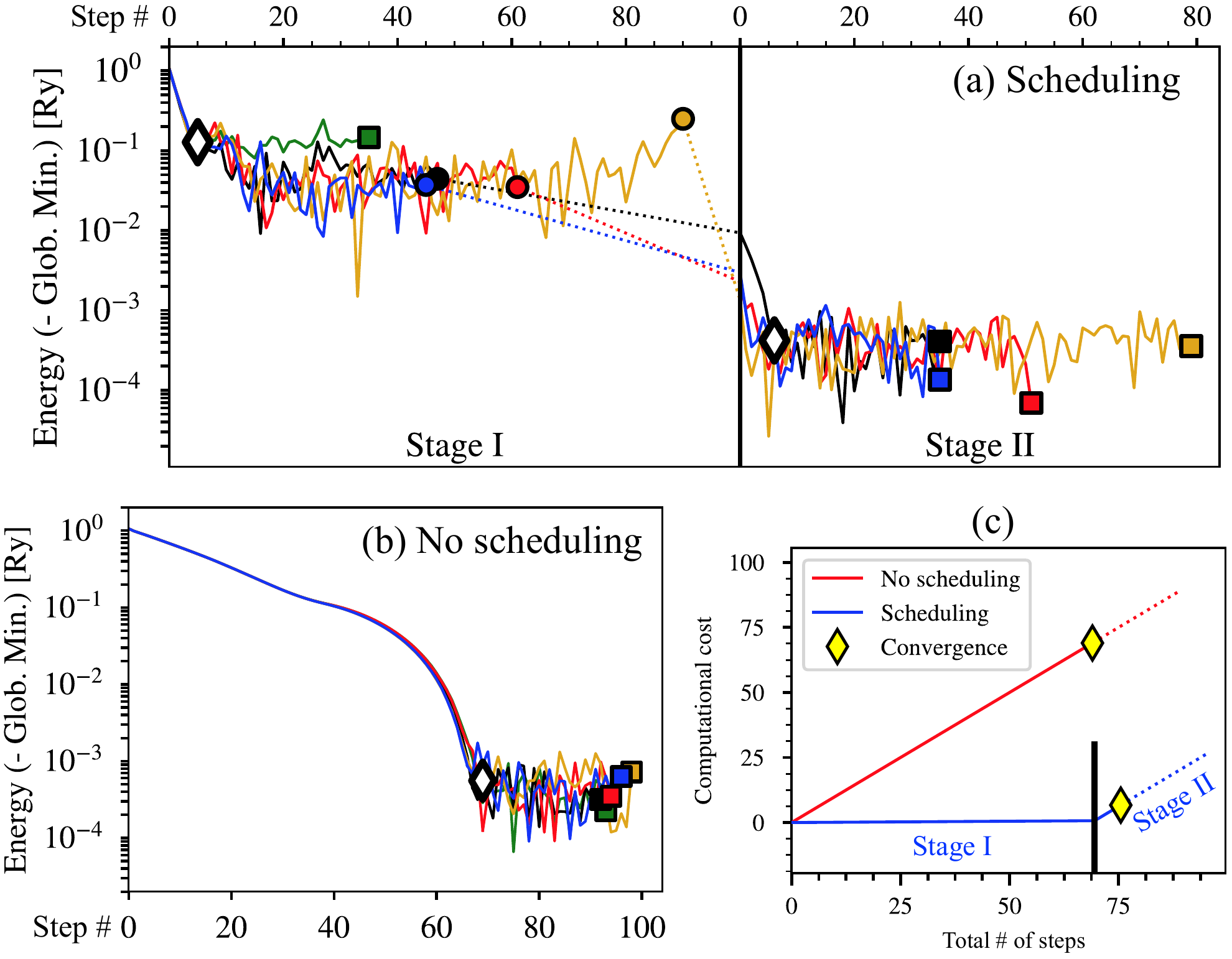}\caption{\label{fig:scheduling-proc}Illustration of SET, and %
the acceleration in optimization efficiency. %
(a) The convergence process of five runs  using FSSD with a 2-stage SET. 
End position of each stage is given by a filled symbol (squares terminated, circles continued). Empty diamond indicates average number of steps before convergence (from posterior analysis).
Four of the five runs are continued in Stage II, after position averaging, indicated by the dotted lines.
(b)  The same as (a) but without SET. 
(c) Comparison of computational costs between (a) and (b).  Total computational cost is shown in 
units of the
time each step takes in (b). 
 Yellow diamonds mark the average
convergence step ($x$-axis) and the average computational cost ($y$-axis) of the five runs.
}
\end{figure}

When the FSSD is applied under the SET approach, a qualitative leap in capability and efficiency is achieved. 
In Fig.~\ref{fig:scheduling-proc}, we illustrate their integration and demonstrate the efficiency gain by their synergy,
using the example of optimization 
in $\mathrm{MoS}_{2}$.
In panel (a), a simple two-stage scheduling is applied in SET. 
The convergence process is shown for five optimization runs.
In each stage, the end of each run is indicated by filled symbols. The automatic script also identifies, after the fact, an  
initial position of convergence, as described in (4) in Sec.~\ref{subsec:SET};
the average of this position in each stage is indicated by the empty diamond. 
A clear lag is seen between the two, leaving a considerable number of steps for position averaging in each run. 
Position averaging ensures that these steps are not wasted but effectively utilized. 
This is reflected by the drastically better initial positions in Stage II than the corresponding end positions in Stage I,
as seen in the lowering of the error in the energy. %
One of the runs (green curve) is discontinued after
Stage I, because it is trapped in a local minimum, as identified by the clustering of the converged positions from all the runs.
In stage II,
the step size and error target are both reduced by a factor of 10.  

Panel (b) in Fig.~\ref{fig:scheduling-proc}  shows the convergence plot without SET. The step size
and error target are fixed at the values used in Stage II above, so that  the same convergence quality is achieved as in (a). 
We see that all five runs converge in this setting. 
From Panel (c), which compares the computational costs between (a) and (b), we see that the two-stage SET procedure resulted in 
a 90\% saving, or ten-fold gain in efficiency in the optimization.

There are two key ingredients in the SET approach:
position averaging at the end of each stage, %
and %
 discrete, staged scheduling instead of adapting the error-bar
and step-size continuously with time.  
In FSSD, a larger step size will generally lead to faster convergence; however, it will result in worse final convergence quality,
because the atomic positions will fluctuate in larger
magnitudes around the minimum. 
\CHANGED{Position (or parameter) averaging helps to dramatically improve the convergence quality FSSD.
The idea of averaging parameters over an optimization trajectory has a long history %
 \cite{Polyak_averaging_1992,Polyak_averaging_1990,Ruppert_averaging_1988} 
and has been applied  in previous structural optimizations in QMC %
(see e.g., Refs.~\cite{Wagner_PRL_2010,Barborini_JCTC_2012,Guareschi_JCTC_2013}).
Our algorithm defines a precise and efficient scheme to apply position averaging retroactively after convergence has been detected.}
It %
allows a wide range of choices for step size, with almost no effect on the convergence
quality, as illustrated in Appendix \ref{sec:position-avg}.
The convergence quality within this range is dictated by the target
error bar size $s$. This makes it more natural to introduce the concept of a separate stage, in which 
we target a smaller error bar (with increased computational cost), and
reduce the step size at the same time to account for the reduced system
scale. 
Comparing to a smooth scheduling procedure, we find this staged scheduling to be efficient, more robust, and resilient to saddle points.

We mention some possible improvements to the SET algorithm over our present implementation. 
We have chosen to reduce %
$s$ and the step size $L$ by the same scale when entering a new stage. 
Around the minimum, %
the optimal step size $L$ is essentially proportional to the distance
$D$ to the minimum, suggesting a choice of $0.1D\sim0.2D$ for  $L$.  
The target error bar $s$ on the force should also be reduced with $D$, but as  illustrated in Appendix \ref{sec:position-avg},
$D$ decreases %
more slowly than $s$.
This indicates
that it would be more optimal to reduce $s$ %
faster than $L$. %
A related point is how much to reduce $s$ in each stage of the scheduling. 
If the choice is too aggressive, a large reduction in $L$ would be required to reach convergence, which in turn would require a large number of steps, 
hence large computational cost. 
If a very small reduction of $s$ is used, %
a large number of stages will be needed,
which is less optimal since there is a threshold of steps to identify convergence in each stage.
Our empirical choice of  $\sim\times 10$ is based 
on the balance of these two extremes.

It is worth emphasizing that %
SET can be employed in combination with other algorithms.
 For example.
we find that position averaging
can improve the convergence quality in 
(by-norm) RMSProp by a similar extent %
to what is seen with FSSD. 
The $\mathrm{RMSProp\times SET}$ approach, although slightly slower than $\mathrm{FSSD\times SET}$ in the examples we studied, would provide
 more freedom in the choice of the step size, 
 as RMSProp allows
for small auto-adaptions.

\CHANGED{
Finally, we comment on the computational cost and scaling of the overall FSSD$\times$SET algorithm. Under optimal step size and error bar sequence choices, the number of steps taken within each stage is roughly the same. The last stage dominates the computational cost associated with the force or gradient computation (Fig.~\ref{fig:scheduling-proc} (c)), and computational cost per step is proportional to the inverse square of error bar. The overall computational cost is then proportional to the inverse square of target precision.
}

\section{A realistic application in AFQMC \label{sec:results}}

We next apply our algorithm to perform a fully \textit{ab initio} quantum many-body geometry optimization in Si. 
Recent progress has made possible the direct computation of
atomic forces and stresses by plane-wave auxiliary-field
quantum Monte Carlo (PW-AFQMC)  \cite{SC_fs_paper}.
Employing this framework, we study the pressure-induced structure phase transition from the insulating diamond phase to the semi-metallic 
$\beta$-tin phase. The detailed setup of this system is given in
Appendix \ref{sec:structures}. 

Figure~\ref{fig:Si-scheduling-QMC} shows the energy difference and
Euclidean distance relative to the target 
 $\beta$-tin
structure in each step during the geometry optimization process. The
run is divided into two stages.
 In stage I, our convergence analysis  identified convergence at
step 26. (See Sec.~\ref{subsec:sched-discuss}.)
Atom positions are accumulated and averaged
starting from this step, yielding a lower and more stable Euclidean
distance curve. This averaged position is taken to be the starting
point $\mathbf{x}_{0}$ for the second stage.
In the second stage 
the statistical error and the step-size are reduced to $2/7$ of the first stage.
The optimization quickly converges
and approaches the correct $\beta$-tin structure. 
The total energy in the final structure is consistent with the ground-state energy computed by AFQMC at 
the ideal $\beta$-tin structure, and the final structure is in agreement with 
the ideal structure %
within our targeted \CHANGED{precision} %
(Euclidean distance of $\sim0.1$\,Bohr).

\begin{figure}
\includegraphics[width=1\linewidth]{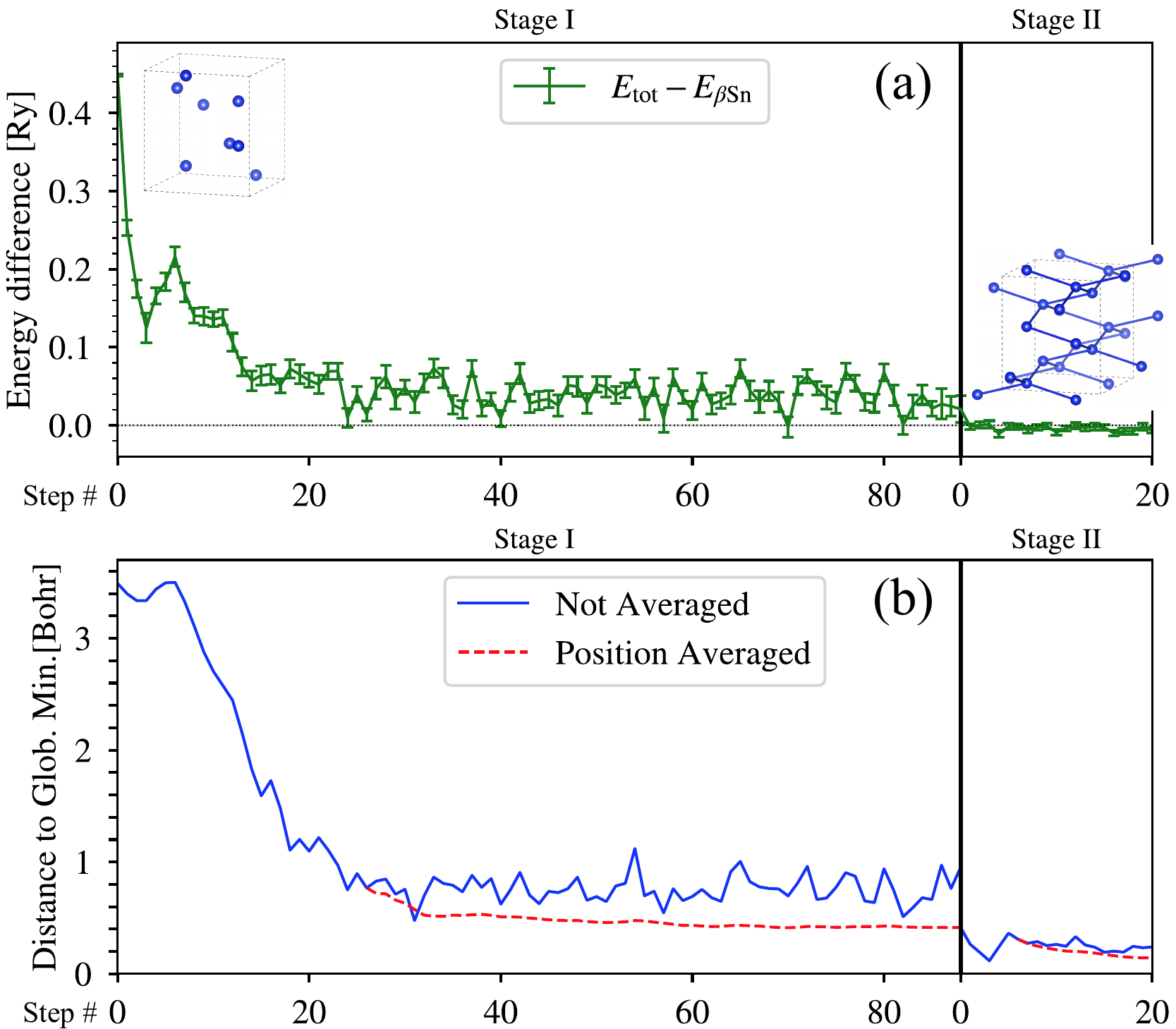}\caption{\label{fig:Si-scheduling-QMC} A direct PW-AFQMC geometry optimization 
with the $\mathrm{FSSD\times SET}$ algorithm. 
Panel (a) shows the total energy relative to the target $\beta$-tin
structure versus the optimization step. %
Error bars are outputted from PW-AFQMC, and reflect the statistical uncertainty of the total energy.
Panel (b) shows the Euclidean distance to the
expected $\beta$-tin structure.
Position averaging result is indicated by the red dashed line, starting when convergence was identified in either stage. %
}
\end{figure}

\section{A new structure in Si \label{sec:results2}}

A (meta)stable orthorhombic structure %
in Si  was discovered 
accidentally in our study.
In this section we present this structure, which to our knowledge was not known.
The new structure emerged in tests of our algorithm for full geometry optimization in solids allowing both the atomic positions and the lattice structure 
to relax.

To apply our algorithm to a full geometry-lattice optimization, we combine the atomic position vectors
and strain tensor into a single generalized position %
\begin{equation}
\mathcal{X}=(\mathbf{x};\{\epsilon_{11},\epsilon_{22},\epsilon_{33},\epsilon_{12},\epsilon_{13},\epsilon_{23}\})\,,
\end{equation}
and the interatomic forces %
and stress tensor into a single gradient %
\begin{equation}
\mathcal{F}=(\mathbf{F};\Omega\,\{\mathbf{\sigma}_{11},\sigma_{22},\sigma_{33},\mathbf{\sigma}_{12},\sigma_{13},\sigma_{23}\})\,,
\end{equation}
such that $\mathcal{F}=-\partial E(\mathcal{X})/\partial\mathcal{X}$ as before.
The cell volume  $\Omega$  appears above, which
is included in the definition of the stress tensor: 
$\sigma_{ij}=-(1/\Omega)(\partial E/\partial \epsilon_{ij})$ \cite{Martin_ES_2020}.
Care must be taken with metrics, e.g. the
step size $L$ in the algorithm should be defined as
\begin{equation}
L=\sqrt{|\Delta\mathbf{x}|^{2}+\sum_{i\leq j}(\nu^{-1}\Delta\epsilon_{ij})^{2}}\,,
\end{equation}
where $\nu$ has the dimension of inverse length. 

\begin{figure}
\includegraphics[width=1\linewidth]{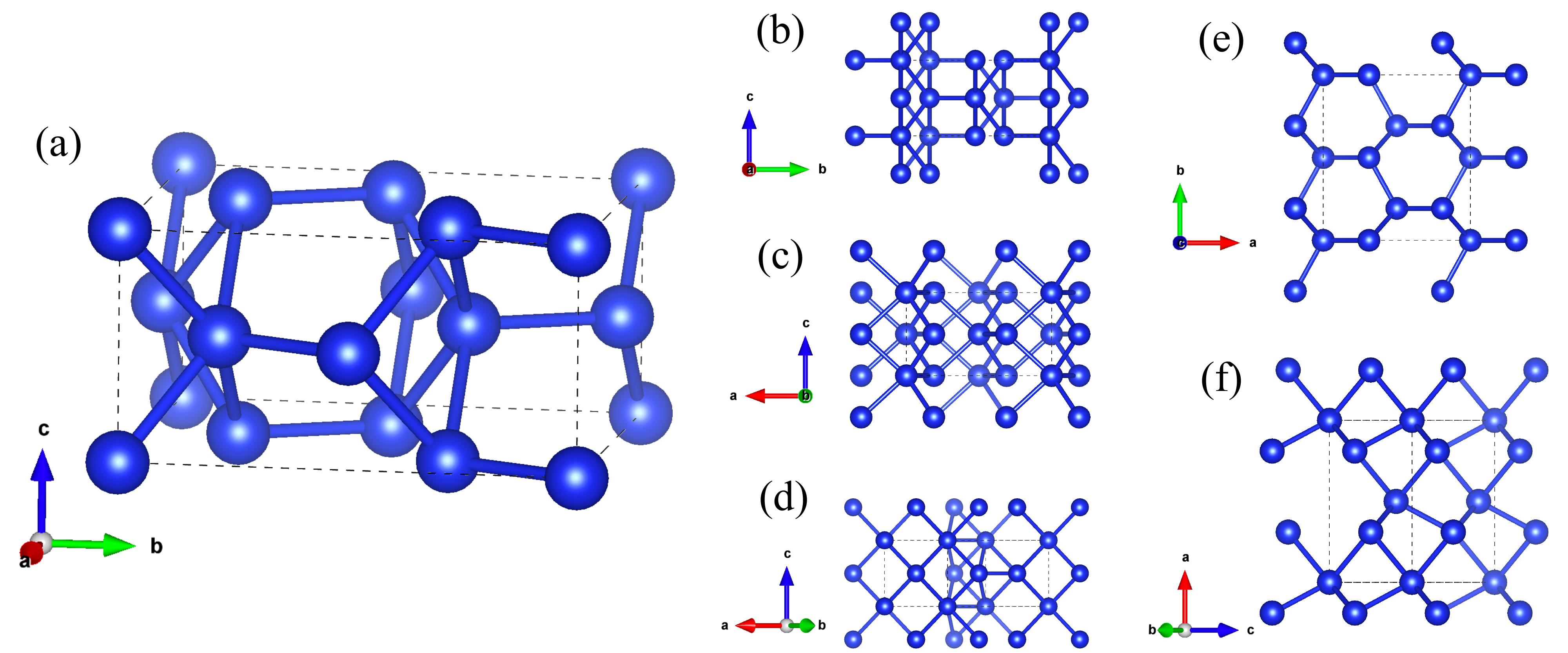}\caption{\label{fig:Cmca-structure}A new structure in Silicon.
Our optimization algorithm identified a  \textit{$Cmca$} structure which is (meta-)stable in silicon. 
This is an 8-atom conventional orthorhombic supercell, in which
$\alpha=\beta=\gamma=90\lyxmathsym{\protect\textdegree}$, as illustrated in (a). Panels (b), (c), (d), (e), and (f) show the structure 
from different perspectives, as indicated by the orientation in the lower-left corner in each.  
In all the plots, bonds
are added for neighboring Si atoms within $<2.6\mathrm{\lyxmathsym{\protect\AA}}$
(plots are generated with the VESTA software \cite{Koichi_VESTA_2008}). 
}
\end{figure}

An additional role of  $\nu$ is to tune the optimization procedure, as it controls the relative step size for optimizing the atomic positions 
versus the overall lattice structure. Different choices thus can result in different optimization trajectories. %
As we describe in detail in Appendix~\ref{app:tab-new-struc}, there is considerable sensitivity of the optimized structure (local minimum) 
with respect to the choice of $\nu$, as well as an interplay with the particular stochastic realization of the optimization trajectory. In general 
this would seem to be an additional  disadvantage of optimization in the presence of stochastic gradients. However, it provides a natural realization of 
statistical sampling of the landscape which could  broaden the search in the optimization. It is this feature that
 lead to the surprise discovery of the new structure shown in Fig.~\ref{fig:Cmca-structure}.

The structure identified has an energy $+0.312$ eV/atom higher than that of the ground state in the diamond structure, determined by DFT PBE
calculations.
We have verified that it is a meta-stable state. We sampled 
5,000 different perturbations around the structure with randomly displacements 
in both atomic positions and lattice distortions,
confirming that all resulted in higher energy.
This was followed by the computation of a Hessian matrix, 
by fitting the total energy with a second-order Taylor expansion, 
which was found to be positive with respect to all geometrical degrees of freedom.

\section{Conclusion}
\label{sec:conclusion}

We have proposed a new structural optimization algorithm to work with stochastic forces and gradients. 
The presence of statistical error bars in the gradients is a common characteristic in many  quantum 
many-body computations. We find that existing optimization algorithms all experience significant difficulties in such 
situations. This is a fundamental problem %
whose importance is magnified by both the growing demand for the higher predictive power and
the generally high cost of \textit{ab initio} many-body calculations.
Our algorithm addresses this problem by the combination of a fixed-step steepest descent
and a staged error scheduling with position averaging. 
The algorithm is simple and straightforward to implement.
It out-performs standard optimization methods used in structural optimization, 
as well as several machine-learning methods, in our extensive analysis
performing realistic geometry optimizations in solids.
The algorithm is then applied in an actual \textit{ab initio} many-body computation,
using plane-wave auxiliary-field quantum Monte Carlo to realize a full structural optimization.
This marks a milestone in the optimization of a quantum solid
using systematically accurate many-body forces beyond DFT.  

The optimization algorithm can be applied to atomic position and lattice structure 
optimizations, as well as a full geometry optimization combining both.
We demonstrated the combined approach for a full geometry optimization, which resulted in the 
discovery of a new structure in Si. 
Furthermore, we illustrated that the presence of statistical noise sometimes creates new opportunities in the optimization.
This can be in the form of tuning the 
\CHANGED{target statistical error}
to minimize the computational cost,
or exploiting the noise to alter the optimization paths and expand the scope of the search,
in the spirit of simulated annealing.

In addition to geometry optimization, the algorithm can potentially be applied to other problems in which the gradients contain stochastic noise. 
The two components of the algorithm can be applied independently or combined with other methods. Insights from them can also stimulate 
further developments.
With the intense effort in many-body method development to improve the predictive power in materials discovery, more efficient optimization methods which handle 
and take advantage of the stochastic nature of the gradients will undoubtedly find ever-increasing applications.

\section*{Code availablity}
Code of FSSD$\times$SET is available at \url{https://github.com/schen24wm/fssd-set}.

\begin{acknowledgments}
We thank B. Busemeyer, M. Lindsey, F. Ma, M. A. Morales, M. Motta,  A. Sengupta, S. Sorella, and Y. Yang for helpful discussions. 
S.C. would like to
thank the Center for Computational Quantum Physics, Flatiron Institute
for support and hospitality. We also acknowledge support from the
U.S. Department of Energy (DOE) under grant DE-SC0001303. The authors
thank William \& Mary Research Computing and Flatiron Institute Scientific
Computing Center for computational resources and technical support.
The Flatiron Institute is a division of the Simons Foundation.
\end{acknowledgments}

\bibliography{GeoOpt_Paper_v1.3.3}

\pagebreak
\widetext
\begin{center}
\textbf{\large Supplemental Materials}
\end{center}

\appendix

\section{The Effect of Noise in Line-search Methods\label{sec:fragile-line-search}}

\begin{figure}
\includegraphics[width=0.5\linewidth]{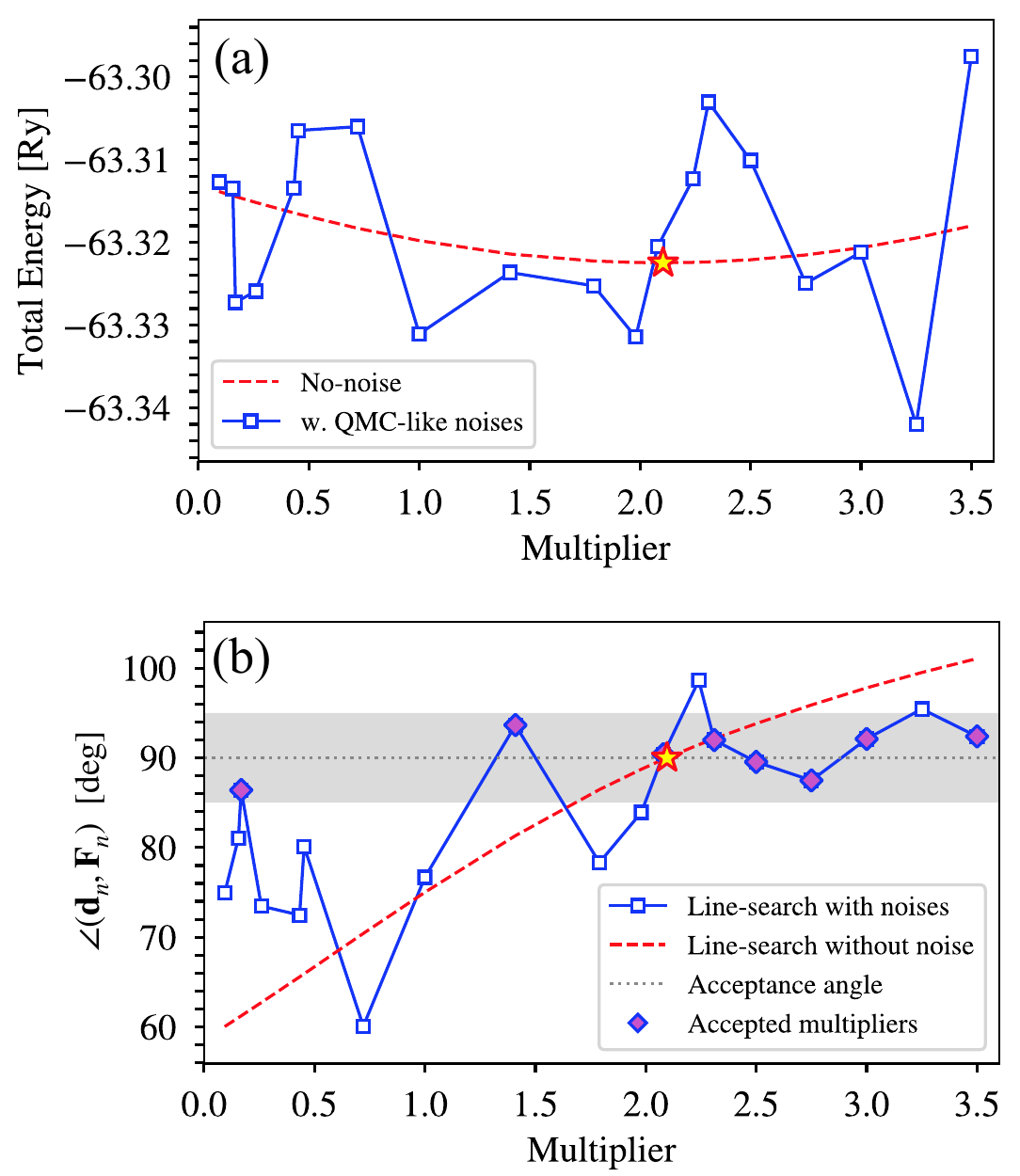}\caption{\label{fig:line-search-problem}%
Illustration of the effect of statistical noise on %
line-search algorithms. 
Panel (a) shows the total energies, %
with the red dashed line indicating the exact energy values (i.e., when the statistical error is driven to zero)
and solid blue shows the noisy energy values (i.e., with  %
statistical noise of the targeted magnitude included). The yellow star indicates the %
``correct'' energy minimum. Plot (b) shows how the angle between
the search direction and the resulting force, a metric for efficient automatic
line search, changes with different displacement multipliers $\Delta x/|\mathbf{F}|$.
An energy minimum on the search direction corresponds to a $90\lyxmathsym{\protect\textdegree}$
angle.
The red dashed line
is for line-search with noise-free forces, 
which gives one clear candidate
at a multiplier of $\sim$2.10 (yellow star). The blue solid line
is for noisy forces and shows multiple candidates (purple diamonds)
falling into the $5\lyxmathsym{\protect\textdegree}$ tolerance region
(light gray).}
\end{figure}

Two of the most common geometry optimization methods are the steepest
descent \cite{Debye_SD_1909} and conjugated gradient
\cite{Magnus_CG_1952,Shewchuk_CG_1994,FletcherReeves_CG_1964,PolakRibiere_CG_1969},
both of which use line-search
\cite{RobbinsMonro_LS_1951,Armijo_LS_1966,Wolfe1,Wolfe2,Bertsekas_SIAMOpt_2006,Bertsekas_LS_2016}
as a building block. Such methods are fragile in %
the presence of %
noisy forces. Here we use steepest descent as an example. 
The search direction is given
by $\mathbf{d}_{n}=-\nabla E_{n-1}=\mathbf{F}_{n-1}$. %
The next step position is chosen along this direction, at or near %
the energy
minimum. This position is found by either manually choosing a few
points to fit a curve, or by automatically selecting a few points
until a criteria is reached, following a %
line-search algorithm.
An example is given in Fig.~\ref{fig:line-search-problem}, using the Newtonian line-search algorithm.
The minimum is the %
point $\angle (\mathbf{d}_{n},\mathbf{F}_{n})\approx90\lyxmathsym{\textdegree}$,
when the force is perpendicular to the search direction $\mathbf{d}_{n}$.

This line-search method works well %
for  forces without noise,
but can run into difficulty %
with noisy forces. %
Noisy forces can cause multiple candidates for
$\angle(\mathbf{d}_{n},\mathbf{F}_{n}) \approx 90\lyxmathsym{\textdegree}$
to appear. 
Since line-search algorithms usually set a tolerance 
to avoid excessive searching,
this can result in the algorithm stopping 
at an undesired multiplier when the threshold is reached. In the example of Fig.~\ref{fig:line-search-problem}(b), we set a tolerance of  $5\lyxmathsym{\textdegree}$, 
and the run stops at a multiplier of $\sim$0.2, with an angle of
$86.4\lyxmathsym{\textdegree}$. This is far %
from the
real minimum, which is at a multiplier of $\sim$2.0. 
Although %
the energy fluctuations are not directly involved in the runs, %
the statistical noise in forces is directly inherited
from the energies, which are shown %
in Fig.~\ref{fig:line-search-problem}(a). 
The same difficulty is manifested from either the perspective of the forces or the total energy.

\section{Setup and Details of the Geometry Optimization Examples\label{sec:structures}}

Two realistic optimization problems from quantum solids are used as test cases in this paper. 
The first is a phase transition
in silicon under pressure \cite{Wirawan_PRB_2009}. The second involves phases in a two-dimensional material, %
monolayer molybdenum disulfide
($\mathrm{MoS_{2}}$). %

\textit{Silicon phase transition.} We explore the phase transition
between the diamond (Si-I) and  beta-tin
structure (Si-II). The parameters of the diamond-structure
is taken from experiment: the primitive cell is an face-centered cubic
(FCC) with lattice constant of 10.263 Bohr \cite{Kittel_SSPhys_1996}. The beta-tin structure
is only stable under high pressure, with an experimental lattice constant
$a$ of 8.82 Bohr and $c/a$ of 0.550 under 11.7 GPa \cite{McMahon_PRB_1994}. 
At zero pressure, DFT (all-electron LAPW) predicts the equilibrium beta-tin structure with a 
lattice constant of about 8.988 Bohr and $c/a$
of 0.552 \cite{Wirawan_PRB_2009}. 

\begin{figure}
\includegraphics[width=0.5\linewidth]{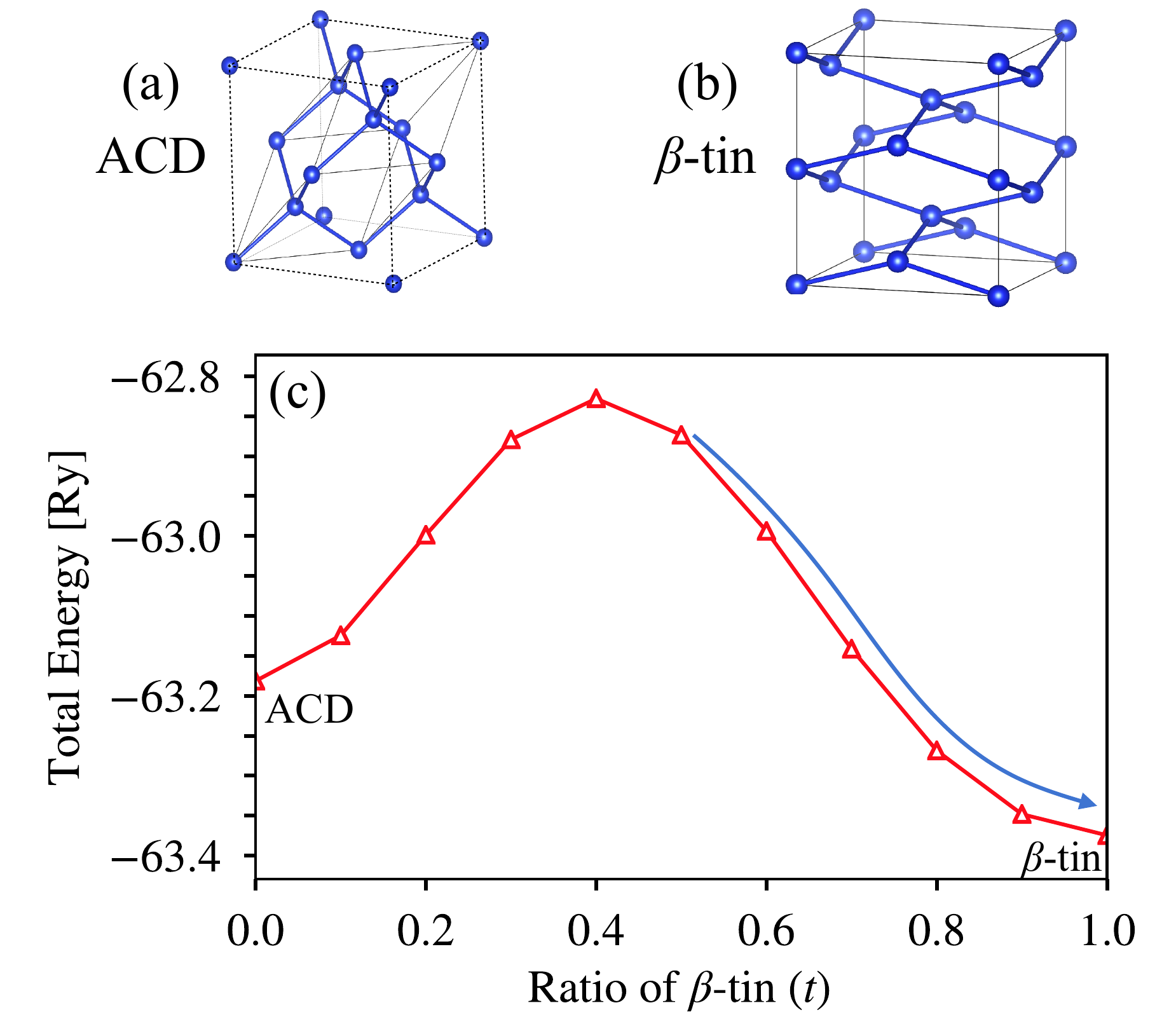}\caption{\label{fig:Si-setup}Setup of the Si phase transition. (a) and (b)
shows the 8-atom supercell for anisotropically compressed diamond
(ACD) structure Si and $\beta$-tin structure Si (Si-II), respectively.
ACD mimics the Si diamond structure (Si-I), which is cubic and has different lattice constants from Si-II.
(c) shows the total energy vs.~the structure  %
between ACD and
$\beta$-tin , where $(1-t):t$ is the mixing ratio of ACD and $\beta$-tin
in the structure. Our starting structure is $t=0.5$,
and the blue arrow indicates the optimization (of atomic positions only). 
}
\end{figure}

In the force-only optimization, we consider an 
``anistropically compressed diamond (ACD)'' structure: the diamond
structure is compressed from the experimental cubic cell to the beta-tin
cell, with the $x$ and $y$ directions compressed
to 8.988 Bohr and $z$ direction to 9.922 Bohr ($c/a=1.104$).
This is a meta-stable structure (local minimum), mimicking  the diamond structure 
within the choice of supercell
size and shape. 
The optimization starts from an equal mixture
of ACD and beta-tin. This ``50:50 mix'' is
the middle point on the closest route that moves all atoms
in ACD to their corresponding beta-tin position, considering
all possible atom swaps, crystal symmetries, and translation symmetry.
FIG. \ref{fig:Si-setup} (c) shows a plot of the total %
energy along this route. 
The middle point (50:50 mix) is close to the
energy barrier but on the beta-tin side. 

For our full geometry optimization, we start from ACD as well, but with the experimental lattice constants at the phase transition ($a=8.82$ Bohr, $c/a=1.100$).

\begin{figure}
\includegraphics[width=0.5\linewidth]{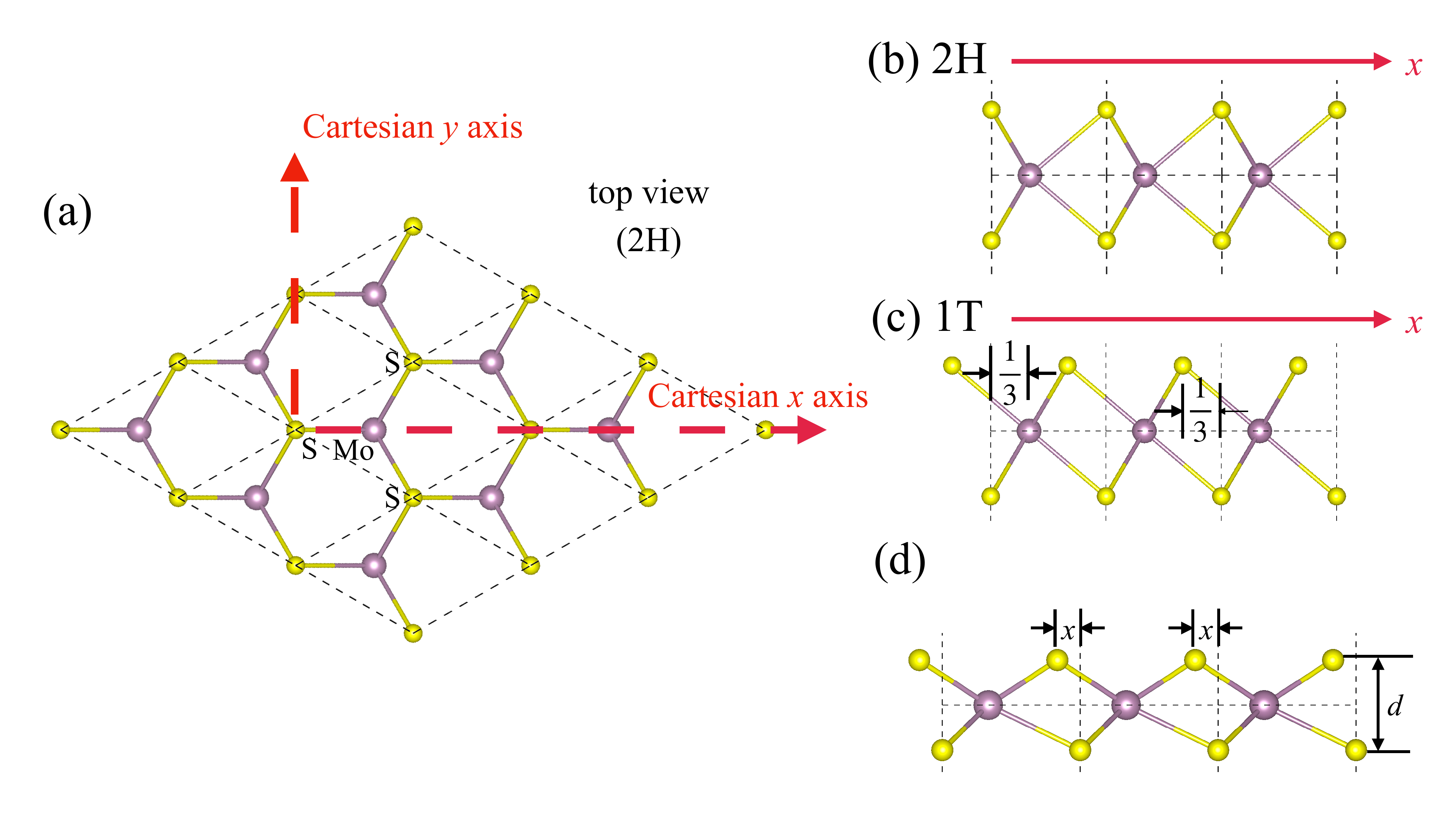}\caption{\label{fig:MoS2-setup}Setup of the $\mathrm{MoS}_{2}$ geometry optimization
problem. (a) shows the top view of the global minmium 2H %
structure. Each S atom in the plot represents two S atoms overlapping
in the \textit{z} direction. (b) shows the side view of the 2H structure,
which shows the correct \textit{z}-positions of the two S atoms. (c)
shows the most common local minimum 1T structure. The top S atom
is displaced from its 2H position by 1/3 of the lattice interval along the
\textit{x} direction. (d) illustrates the parametrization we use to distinguish the two structures. 
There are seven (7) other degrees of freedom to fully specify the geometry of each phase, which 
are optimized simultaneously with $x$ and $d$.
}
\end{figure}

\textit{$\mathrm{MoS}_{2}$ monolayer.} This is a two-dimensional
system with a finite layer thickness in the third direction. Simulations
of such a system is done with a large \textit{z} axis lattice constant
(36.12 Bohr in our case). 
There are two stable
sulfur atom alignments \cite{He_MoS2_2016}: one  is the
2H global minimum where the two S atoms are stacked together at the
top view, and the other is the 1T local minimum 
where the Mo and each of the two S atoms sit at the 3 possible hexagonal sites (see Fig.~\ref{fig:MoS2-setup}).
The thickness of the layer,  the
S-S atom %
distance in \textit{z}, is tunable and not constrained
by any symmetry requirements. 
Thus the two phases are characterized by two controllable
parameters: $x$ which gives the S-S alignment mismatch, and $d$
which gives the layer thickness. In our definition, $x=0$ gives 2H,
and $x=-\frac{1}{3}$ gives 1T. We work %
with the 3-atom primitive
cell, starting from a system %
with layer thickness $d$ compressed
to 1.8 Angstrom and one of the S atom moved to a 50:50 mix of the
2H and 1T structure ($x=-\frac{1}{6}$). Together with $x$ and $d$ are seven (7) additional 
free parameters to be optimized which define the structure of the system.

\section{Additional Discussion on SET \label{sec:position-avg}}

\begin{figure}
\includegraphics[width=0.5\linewidth]{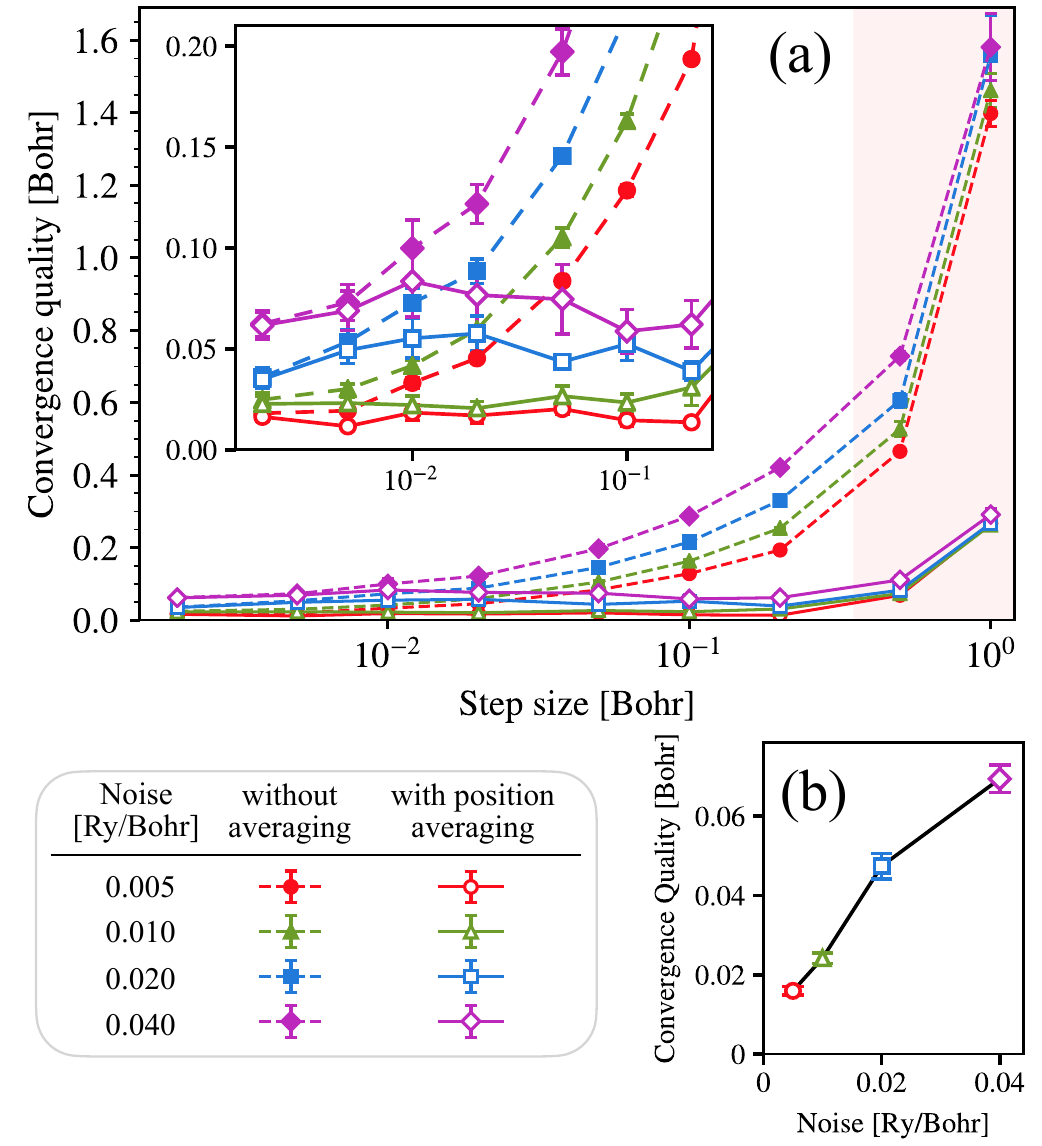}\caption{\label{fig:convergence-step-error} Analysis of the convergence quality
vs.~step size and statistical error (noise) size.
In plot (a), the dashed lines show the convergence quality
without position averaging, while the solid lines show the convergence
quality after position averaging.
Error bars show the standard error of 6 runs with random noises.
Region with a \textit{white} background
is the region where position averaging will remove the dependence
on step sizes in the convergence quality. Region with a \textit{red}
background marks where this dependency can no longer be fully removed,
indicating the step size is too large. The inset provides a zoomed-in
view of the white region. Plot (b) shows the convergence quality vs.~noise size after position averaging. For each noise size, the convergence
quality shown in this plot is computed from the average of the position-averaged
convergence qualities of all step-sizes in the inset of plot (a).
Error bars show the standard error of these 7 step-sizes.
The system is  $\mathrm{MoS}_{2}$ and 
convergence quality is defined by the Euclidean distance (see  Appendix \ref{sec:conv-analysis-algo})  
from the correct minimum.
}
\end{figure}

We study the relation between the step size,
the 
\CHANGED{target statistical error}
on the gradient,
and the convergence speed. 
A larger step size means faster convergence
as long as the step size is not too large to blur the difference between different
minima. However, as shown in Fig. \ref{fig:convergence-step-error}
(a), large step sizes can result in a worse final convergence quality,
due to %
larger  fluctuations in the atomic positions
around the correct minimum. By introducing position
averaging, %
we can %
mitigate the fluctuation so that 
there is almost no effect on the convergence
quality within a wide range of step size choices (white-background
region in the plot), and the final convergence quality %
only
depends on the error bar size (Fig. \ref{fig:convergence-step-error}
(b)). At this point, if a higher precision is still desired, we should
increase the computational cost to target a smaller error bar, and
reduce the step size at the same time to account for the reduced system
scale.

Perhaps a more natural and intuitive approach to the %
scheduling
is to tune the error bar (and step-size)  %
 continuously. %
However, the entanglement of the statistical noise and the effect of retardation in the search process makes this less straightforward.
For example, if the optimization process moves through a flat region (saddle point) and then re-enters a fast convergence phase, a smooth 
control of the target error bar or step size could lead to a significant reduction in efficiency. 

The SET algorithm,  
instead of %
more sophisticated techniques (e.g. P-controller
\cite{Wadia_OPT2021}), %
devises a simple
solution by dividing the runs into stages, in each of which the error
bar size and step size are kept constant. Using stages does
not remove the retardation effect mentioned above; however, by using an automatic convergence
identification algorithm and requiring a relatively long convergence
phase, we can now identify a convergence with confidence, albeit
at a later time.
Combining with FSSD and position
averaging allows a sampling around the minimum in a Monte Carlo sense, and avoids ``wasting''
the extra steps after %
convergence, as illustrated in Fig.~\ref{fig:scheduling-proc}.
We find that this approach makes for a simple, and more robust and efficient algorithm which outperformed all our attempts at continuous scheduling.

\section{Parameter Choices in the Optimization Methods \label{sec:method-params}}

The optimization methods employed in this work all have some
free parameters or variations. We have not attempted to perform the most detailed optimization of 
these parameters. The following describes our choices. %

\begin{figure}
\includegraphics[width=0.5\linewidth]{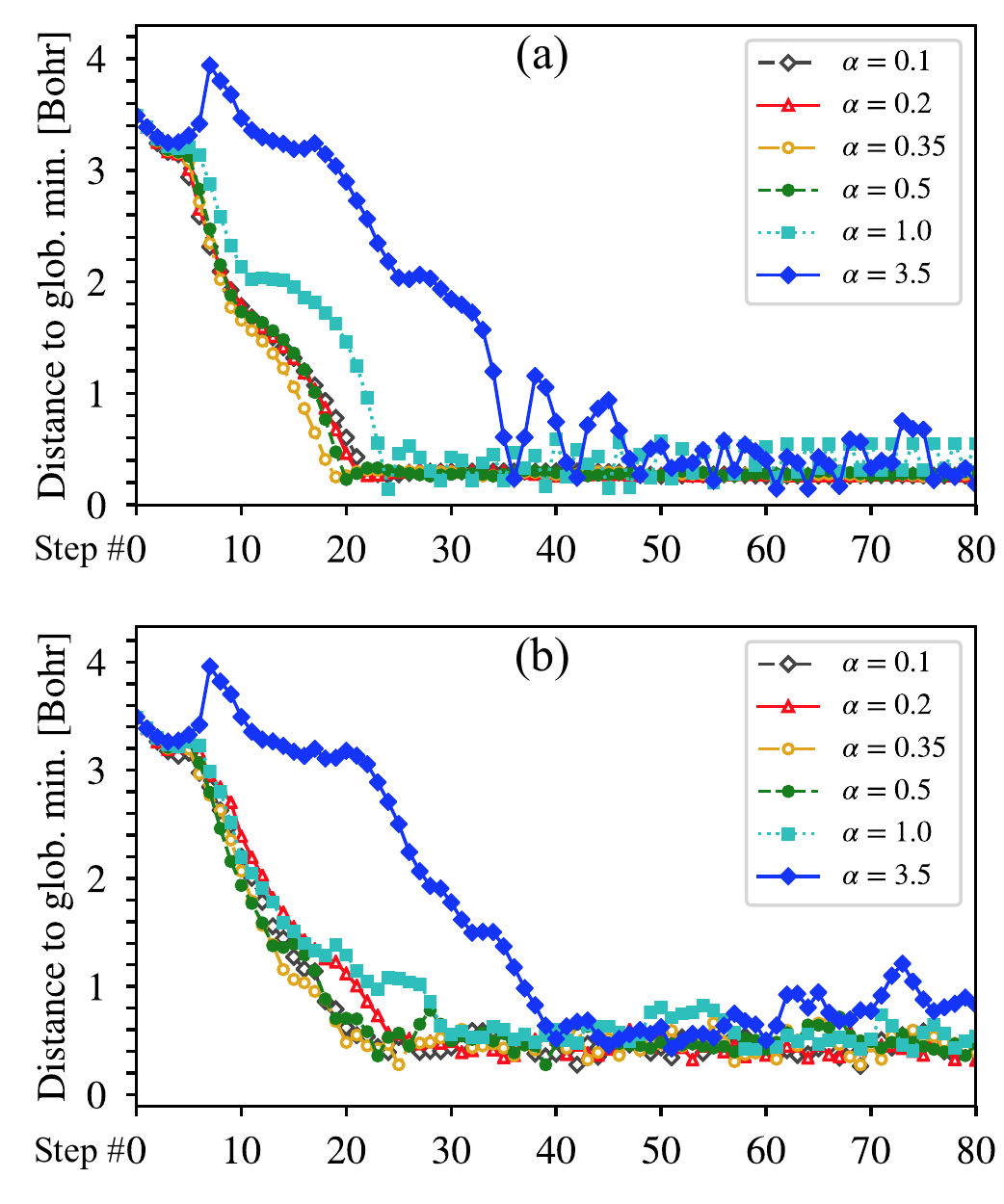}
\caption{\label{fig:mixing-param-research}Convergence analysis of FSSD versus the choice of mixing parameter $\alpha$.
(a) shows %
6 parameter choices,  
with no noise. (b) shows the runs with the same parameters but with synthetic noise introduced.  %
The setup is Si phase transition (see Appendix \ref{sec:structures})
and a fixed step size = 0.5 Bohr is used.
Distance to the global minimum is measured as  the symmetrical Euclidean distance, as defined
in Appendix \ref{sec:conv-analysis-algo}.}
\end{figure}

FSSD uses a step size of 0.5 Bohr for Fig. \ref{fig:Si-compareLSML},
0.3 Bohr for Fig. \ref{fig:MoS2-compareML}, 0.5 Bohr for Fig. \ref{fig:scheduling-proc}
(stage I), and 0.7 Bohr for Fig. \ref{fig:Si-scheduling-QMC} (stage
I). A mixing parameter of $\alpha=1/e$ is used throughout the work. 
Our tests show 
that a value between $0.35\sim0.5$ yields %
good convergence speed and final convergence accuracy (Fig.~\ref{fig:mixing-param-research}).

Steepest-descent and conjugate-gradient are based on a Newtonian line-search
that finds the root of $<\mathbf{d}_{n},\mathbf{F}_{n}>=90\lyxmathsym{\textdegree}$,
as in Appendix \ref{sec:fragile-line-search}. Conjugate gradient
uses the Polak-Ribi\'ere formula and restarts every 5 steps. 
\CHANGED{Note that for noisy PES, specially designed methods \cite{Schraudolph_CG_2003} can result in better performance for conjugate gradient. 
This was not pursued here, since the required automatic differentiation is not always available in the many-body computations with which the 
optimization algorithm is  expected to couple.}

Machine-learning %
based algorithms discussed in this work have
an ``element-wise'' version and a ``by-norm'' version. The version
in the original literature of RMSProp, AdaDelta, and Adam is ``element-wise''
\cite{Tieleman_RMSProp_2012,Zeiler_Adadelta_2012,KingmaBa_Adam_2014};
for example, the RMSProp algorithm is
\begin{equation}
\langle F^{2}\rangle_{n}=0.9\langle F^{2}\rangle_{n-1}+0.1F_{n}^{2}\,,
\end{equation}
\begin{equation}
x_{n+1}=x_{n}+\frac{\eta}{\sqrt{\langle F^{2}\rangle_{n}+\epsilon}}F_{n}\,,
\end{equation}
where $x_{n}$ is the atom position at step $n$, $F_{n}$
is the force computed at position $x_{n}$, $\eta$ is a fixed learning
rate, and $\epsilon$ is a  small number to prevent singularity.
$\langle F^{2}\rangle$ is a ``historical average'' of all squared
forces. This original ``element-wise'' algorithm treats each dimension
separately, and $x_{n}$, $F_{n}$, $F^{2}$, and $\langle F^{2}\rangle$
are all vectors. A variant of this algorithm, %
which we call the ``by-norm'' algorithm, is
given by replacing the equations %
above with
\begin{equation}
\langle|\mathbf{F}|^{2}\rangle_{n}=0.9\langle|\mathbf{F}|^{2}\rangle_{n-1}+0.1|\mathbf{F}|_{n}^{2}\,,
\end{equation}
\begin{equation}
\mathbf{x}_{n+1}=\mathbf{x}_{n}+\frac{\eta}{\sqrt{\langle|\mathbf{F}|^{2}\rangle_{n}+\epsilon}}\mathbf{F}_{n}\,,
\end{equation}
where the force is now treated as a whole for all dimensions,
as each dimension receives the same value as the prefactor for the
force. By analogy, the FSSD algorithm we use should be classified
as a ``by-norm'' algorithm.

Our application of the Adadelta algorithm has one small %
modification from its original form
\cite{Zeiler_Adadelta_2012}, which forces $E[\Delta x^{2}]_{0}=0$
and appears to have poor efficiency in our optimizations. Replacing
this value with a finite number gives the algorithm an initial
boost and specifies the initial step size with $\eta=\sqrt{E[\Delta x^{2}]_{0}/(1-\rho)}$.

We used the following parameters for ML based algorithms. %
RMSProp (element-wise)
used the suggested mixing factor $\beta=0.9$, with an initial step
size $\eta=0.2$ Bohr in each dimension for Si (Fig. \ref{fig:Si-compareLSML})
and $\eta=0.1$ Bohr in each dimension for
$\mathrm{MoS_{2}}$ (Fig. \ref{fig:MoS2-compareML}). By-norm RMSProp
used a step size of 0.7 Bohr for $\mathrm{MoS}_{2}$ (Fig. \ref{fig:MoS2-compareML}).
Adadelta used a mixing parameter of $\rho=0.9$, with
an initial step size $\eta=0.4$ Bohr in Si.
Adadelta (element-wise or by-norm) used the same $\eta$ as RMSProp in $\mathrm{MoS}_{2}$.
Adam (element-wise) used the same initial step size as RMSProp for
all systems, and $\beta_{1}=0.9$, $\beta_{2}=0.999$ as suggested
in the original paper. By-norm Adam had a slower performance, hence
we switched the learning rate from 0.7 Bohr/step to 2.0 Bohr/step in
$\mathrm{MoS}_{2}$.

\section{Convergence Analysis Algorithm\label{sec:conv-analysis-algo}}

The Euclidean distance metric 
between two atom-position arrays $\mathbf{x}$
and $\mathbf{x}_{\mathrm{ref}}$  is
\begin{equation}
D(\mathbf{x},\mathbf{x}_{\mathrm{ref}})=|\mathbf{x}-\mathbf{x}_{\mathrm{ref}}|=\sqrt{\sum_{i=1}^{3N_\mathrm{at}}(x_{i}-x_{\mathrm{ref},i})^2}\,,
\end{equation}
where $N_{\mathrm{at}}$ is the number of atoms.
Periodicity, global translations on all $N_\mathrm{at}$ atoms,
and atom permutations are applied to $\mathbf{x}$ 
to minimize
$D(\mathbf{x},\mathbf{x}_{\mathrm{ref}})$. If $\mathbf{x}_{\mathrm{ref}}$
is a known structure with some crystal symmetry, then the crystal
symmetry operations are also applied. This is the case when we measure
``the distance to global minimum'' to compare algorithm efficiencies.
On the other hand, in convergence analysis %
where we have no knowledge of
the final structure, %
only the periodicity, translations, and permutations are used.

Position averaging is performed with similar consideration for symmetries:
position of the last optimization step is chosen as $\mathbf{x}_{\mathrm{ref}}$, 
and symmetry operations including periodicity, translation, and permutation are applied 
to all earlier steps to minimize the distance to $\mathbf{x}_{\mathrm{ref}}$
before the actual averaging takes place.

\begin{figure}
\includegraphics[width=0.5\linewidth]{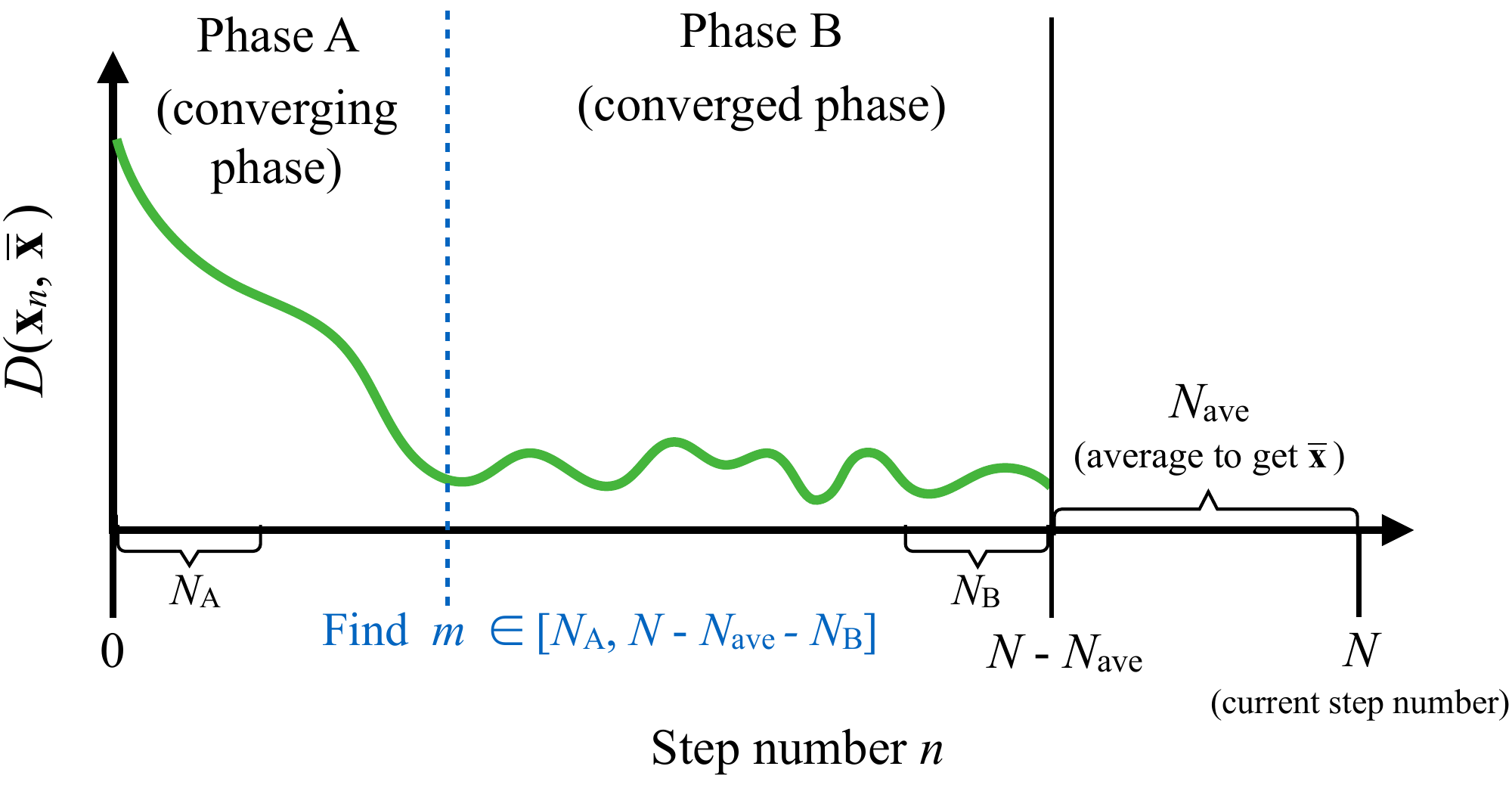}\caption{\label{fig:conv-anal-algo}The convergence analysis algorithm at step $N$. The averaged position of the last $N_\mathrm{ave}$ steps is used as a reference position $\mathbf{x}_\mathrm{ref}$. Euclidean distance between (1) this reference position and (2) atom positions from step 0 to $N-N_\mathrm{ave}$ are computed to construct a one-dimensional distance function (the green curve). The algorithm finds a step number $m \in [N_\mathrm{A},N-N_\mathrm{ave}-N_\mathrm{B}]$ that maximizes the ratio between the standard error of all distance function values in phase A and the this standard error in phase B.
}
\end{figure}

Our convergence analysis algorithm is illustrated in Fig.~\ref{fig:conv-anal-algo}. 
The Euclidean distance metric is used to build %
a one-dimensional distance function of the steps in the optimization
history, with $\mathbf{x}_{\mathrm{ref}}$ being the ``current best
guess.'' At step $N$, $\mathbf{x}_{\mathrm{ref}}$  %
is selected
as the position average $\bar{\mathbf{x}}$ of the last $N_{\mathrm{ave}}$ steps of the
convergence procedure.

We compute $D_{n}=D(\mathbf{x}_{n},\bar{\mathbf{x}})$ for all $0\leq n\leq N-N_{\mathrm{ave}}$,
and then search for a step number $m$
between $N_\mathrm{A}$ and $N-N_\mathrm{ave}-N_\mathrm{B}$ that divides the entire run into two
phases ($N_\mathrm{A},N_\mathrm{B}$ is the ``minimum phase length'' of phase A and B), 
such that the ratio of the standard errors of the distance in the
first phase and that in the second phase is maximized:
\begin{equation}
m=\argmax_{t\in[N_\mathrm{A},N-N_\mathrm{ave}-N_\mathrm{B}]}R_{t}\,,
\end{equation}
\begin{equation}
R_{t}\equiv\frac{\mathrm{stderr}_{n=0}^{t-1}\{D_n\}}{\mathrm{stderr}_{n=t}^{N-N_\mathrm{ave}}\{D_n\}}
\end{equation}
where $\mathrm{stderr}_{n=P}^{Q}(D_{n})$ denotes the standard error
of $\{D_{P},D_{P+1},\ldots,D_{Q}\}$.
Convergence is reached if $R_m>R_\mathrm{th}$, 
where $R_\mathrm{th}$ is a threshold. %

There are a few tunable parameters in this analysis algorithm. By
default we choose $N_\mathrm{A}=N_\mathrm{B}=5$, $N_\mathrm{ave}=10$, and $R_\mathrm{th}=5$.
 These parameters
can be varied. %
Note that low $N_\mathrm{A},N_\mathrm{B},N_\mathrm{ave},R_\mathrm{th}$
might lead to misidentification of the saddle points as equilibrium,
while high parameters can result in %
longer runs.

\section{Sensitivity in the Joint Optimization of Positions and Lattice Structure \label{app:tab-new-struc}}

\begin{table}[b]
\caption{\label{tab:geolatt-convergence} Final structure of the geometry-and-lattice
optimization, for different stress weight $\nu$: a no-noise run and
6 noisy runs are shown per stress weight. ``Diamond'' is the Si-I
structure; ``$\beta$-tin'' is the Si-II structure; \textit{Imma
}\cite{McMahon_PRB_1994} is a transition structure between Si-II and Si-V (simple
hexagonal); \textit{$Cmca$} is a new orthorhombic structure. 
}

\begin{ruledtabular}
\begin{tabular}{c|ccccc}
\multirow{1}{*}{$\nu$ [Bohr$^{-1}$]} & 0.040 & 0.020 & 0.015 & 0.010 & 0.005\tabularnewline
\hline 
no-noise & \textit{$Cmca$} & diamond & diamond & $\beta$-tin & $\beta$-tin\tabularnewline
\hline 
noisy \#1 & diamond & diamond & diamond & \textit{Imma} & \textit{Imma}\tabularnewline
noisy \#2 & diamond & diamond & diamond & \textit{Imma} & \textit{Imma}\tabularnewline
noisy \#3 & diamond & diamond & diamond & \textit{Imma} & \textit{Imma}\tabularnewline
noisy \#4 & \textit{$Cmca$} & diamond & diamond & \textit{Imma} & \textit{Imma}\tabularnewline
noisy \#5 & \textit{$Cmca$} & diamond & \textit{Imma} & \textit{Imma} & \textit{Imma}\tabularnewline
noisy \#6 & \textit{$Cmca$} & \textit{$Cmca$} & \textit{$Cmca$} & \textit{Imma} & \textit{Imma}\tabularnewline
\end{tabular}
\end{ruledtabular}

\end{table}

The optimization result can depend on the stress weight $\nu$.
A good guess of $\nu$ is the ratio of the optimal strain step
size in a stress-only lattice optimization vs.~the optimal atom-position
step size in a force-only geometry optimization, but different $\nu$ can %
be chosen to emphasize the two aspects (atomic positions vs.~lattice structure).
TABLE \ref{tab:geolatt-convergence} shows the result of an FSSD optimization with different choices of $\nu$. We use
DFT, starting from a 50:50 diamond/beta-tin structure at the diamond-to-beta-tin
transitional lattice constant\cite{McMahon_PRB_1994}. 
For smaller $\nu$ %
the lattice structure is optimized more
cautiously and 
the optimization tends to reach the
\textit{Imma} structure \cite{McMahon_PRB_1994}. 
In the presence of noise, the optimization %
does not reach the beta-tin
structure %
($a=b$ and $\Delta=1/4$) unless the step size and statistical error is made very small.
A larger choice of $\nu$ leads to the diamond structure which is
the global minimum at zero pressure. 
An even larger choice of $\nu$ ($\nu=0.04$)
has a large chance of landing on the %
new
orthorhombic %
structure. %
This example also shows that the presence of noise can sometimes help 
find a new structure by adding a small annealing effect.

\section{The \textit{$Cmca$} structure\label{sec:Cmca-structure}}

The new \textit{$Cmca$} structure is shown in FIG. \ref{fig:Cmca-structure}.
It has a space group of \textit{$Cmca$} (No. 64),
and each atom in the cell has a coordinate number of 5.
The lattice vectors for the 8-atom conventional supercell are (in
$\mathrm{\lyxmathsym{\AA}}$):
\[
\mathbf{a}_1=(6.055,0,0);\,\,\mathbf{a}_2=(0,6.817,0);\,\,\mathbf{a}_3=(0,0,3.477)\,.
\]
There are 4 atoms in the primitive cell, whose crystal coordinates
\[
\mathbf{x}_{1}=(0,0,0);\,\,\mathbf{x}_{2}=(\Delta_{x}-\frac{1}{4},\Delta_{y}-\frac{1}{4},0);
\]
\[
\mathbf{x}_{3}=(\frac{1}{2},\Delta_{y}+\frac{1}{4},0);\,\,\mathbf{x}_{4}=(\Delta_{x}+\frac{1}{4},\frac{1}{2},0)
\]
where
\[
\Delta_{x}=0.06056,\Delta_{y}=-0.05685\,.
\]
In the 8-atom supercell, 
a shift of $\Delta\mathbf{x}=(0,1/2,1/2)$ on the 4 atom
coordinates above gives the other 4 atoms in the supercell.

\section{Comparisons of FSSD with other methods: more examples\label{sec:add-example}}

In this section we provide three more structural optimization examples in solids 
where we compare the performance of FSSD with other methods.
The disparate  examples, together with the systems discussed in the main text, form a diverse set representing a wide range of structural optimization 
problems in quantum materials. We use emulator models in which the forces and gradients are computed by DFT (PBE) with synthetic noise added 
(without considering covariance, which we found to be small from our AFQMC calculations of the forces and stress tensors).

\begin{figure}
\includegraphics[width=0.8\linewidth]{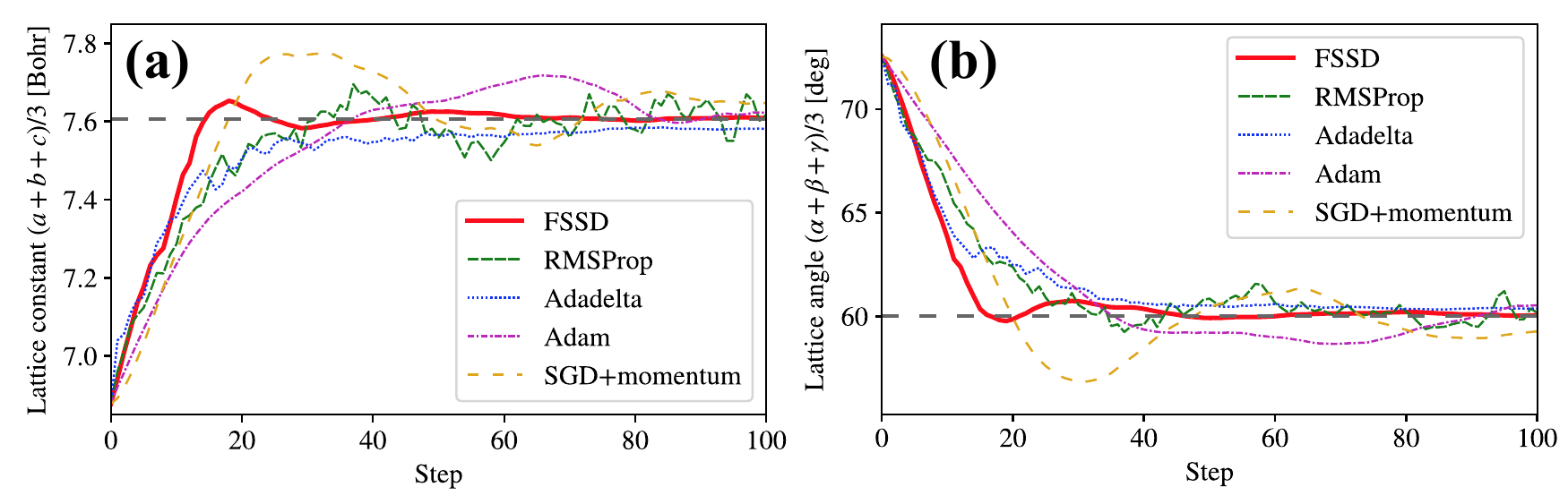}\caption{\label{fig:NaCl-lattopt}\CHANGED{Comparison of FSSD with position averaging
and four ML algorithms in the NaCl lattice optimization problem, showing (a) the average of the three lattice constants and (b) the average of the three lattice angles. Dark gray dashed line indicates the location of the exact minimum.}
}
\end{figure}

The first example is a lattice vector optimization in NaCl, 
where the fractional coordinates of the atoms are fixed at Na $(0,0,0)$ and Cl $(\frac{1}{2},\frac{1}{2},\frac{1}{2})$,
but changes on the lattice vectors (lattice constant and angles) are allowed.
Here the initial structure was chosen to lie roughly at the middle of the rock salt structure 
and the CsCl structure, with the lattice vectors being:
\[
a_0=8.7\,\mathrm{Bohr};\,\,\mathbf{a}_1=a_0(0,\frac{3}{4},\frac{1}{4});\,\,\mathbf{a}_2=a_0(\frac{1}{4},0,\frac{3}{4});\,\,\mathbf{a}_3=a_0(\frac{3}{4},\frac{1}{4},0)\,,
\]
which translates to a lattice constant of $a=b=c\approx6.88$ Bohr and a lattice angle of $\alpha=\beta=\gamma\approx72.54^{\circ}$.
The global minimum is the rock salt structure which is the optimization target. 
FIG.~\ref{fig:NaCl-lattopt} shows how the lattice constants and lattice angles evolve during the optimization process, %
using FSSD %
and four different ML optimization algorithms.

\begin{figure}
\includegraphics[width=1.0\linewidth]{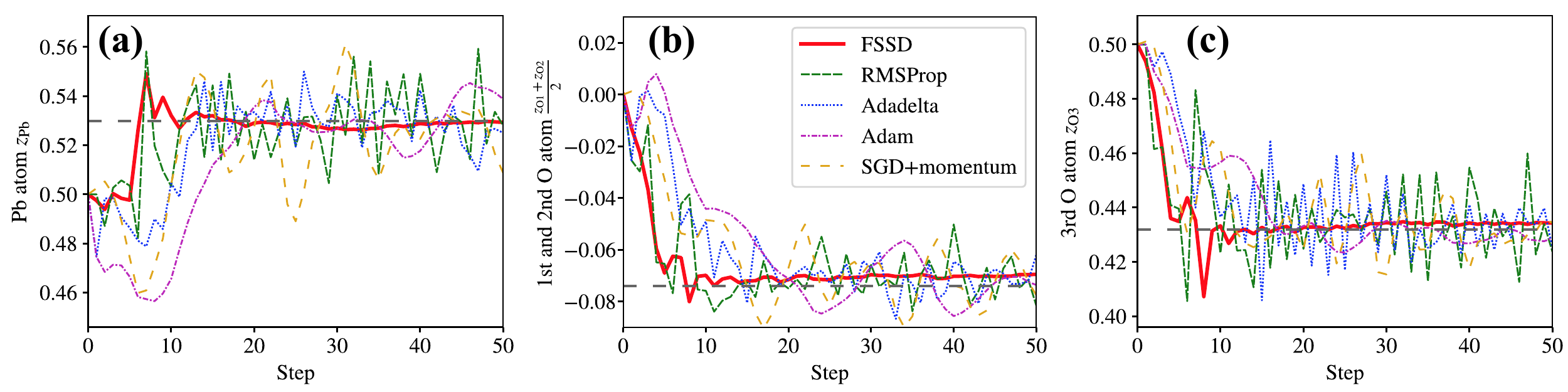}\caption{\label{fig:PbTiO3-opt}\CHANGED{Comparison of FSSD with position averaging 
and four ML algorithms in the PbTiO3 ferroelectric phase transition, showing the z-axis crystal coordinates of (a) the Pb atom, (b) the two O atoms on the short $x$- and $y$-axes (O1 and O2), and (c) the O atom on the long $z$-axis (O3), relative to the Ti atom. %
In (b), the averages of the coordinates of O1 and O2 are shown to reduce the number of curves. %
Dark gray dashed line indicates the location of the exact minimum.}
}
\end{figure}

The second example is in the ferroelectric material $\mathrm{PbTiO_3}$. 
In a tetragonal cell with the experimental lattice constant of $a=7.3775$ Bohr and $c/a=1.0649$  \cite{Kuroiwa_PRL_2001},
the initial centrosymmetric structure is optimized to the ferroelectric phase at the global minimum.
FIG.~\ref{fig:PbTiO3-opt} illustrates this optimization, showing the crystal z-coordinates of the Pb atom and the 3 O atoms,
assuming the Ti atom is fixed at origin.

\begin{figure}
\includegraphics[width=0.6\linewidth]{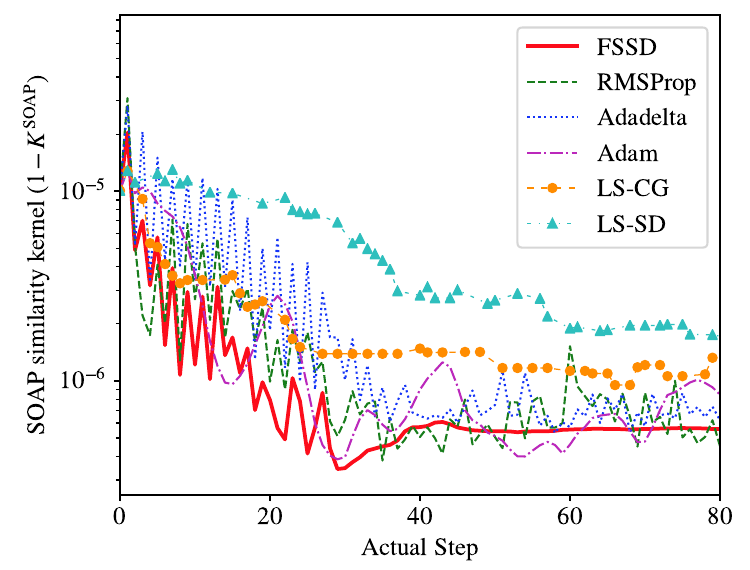}\caption{\label{fig:IceIh-opt}\CHANGED{(a) Comparison of FSSD with position averaging, %
three ML algorithms, and two line search based gradient descent methods (LS-SD: steepest descent, LS-CG: conjugate gradient). The $x$ axis %
shows the number of %
force computations, which is the same as optimization steps in FSSD and the three ML algorithms, but is larger than the optimization steps in the two line-search methods. The $y$ axis shows the SOAP similarity kernel, which measures the  distance between the structure at each step %
and the global minimum.}
}
\end{figure}

The last example is the molecular crystal ice, in a hexagonal lattice setup (ice $I_h$ structure). 
The initial structure has 
one $\mathrm{H}_2\mathrm{O}$ molecule %
displaced by $(0.1,0.1,0)$ in crystal coordinates from equilibrium.
Despite its seeming simplicity, this problem is %
exceptionally difficult because of 
the weak hydrogen bond forces and the resulting high condition number. %
In FIG.~\ref{fig:IceIh-opt}, we show FSSD, three ML algorithms, and two line-search algorithms 
(steepest descent and conjugate gradient). The performance of the algorithms are demonstrated by the similarity distance to the minimum at different optimization steps, measured by the SOAP kernel \cite{Bartok_SOAP_2013} (computed with ASE \cite{ASE1,ASE2} and DScribe \cite{DScribe} packages).

In all three examples, FSSD with position averaging remains one of the fastest method by efficiency, while having also the smallest fluctuations (best accuracy). %
The behaviors of the three ML methods %
(RMSProp, Adadelta, Adam) are also consistent with the observations described in the main text. %
We included gradient descent with momentum (SGD+momentum) in the first two examples, NaCl and $\mathrm{PbTiO}_3$. %
Its behavior is seen to resemble that %
of Adam, characterized by large and slow fluctuations.
This indicates that %
a fixed step size, which is the essential difference between it and FSSD, is critical for the improved performance of FSSD. %

\bibliographystyle{apsrev4-1}

\end{document}